\documentclass[preprint]{aastex}
\usepackage{emulateapj5}
\def\stacksymbols #1#2#3#4{\def\theguybelow{#2}
	\def\verticalposition{\lower#3pt}
	\def\spacingwithinsymbol{\baselineskip0pt\lineskip#4pt}
	\mathrel{\mathpalette\intermediary#1}}
\def\intermediary #1#2{\verticalposition\vbox{\spacingwithinsymbol
	\everycr={}\tabskip0pt
	\halign{$\mathsurround0pt#1\hfil##\hfil$\crcr#2\crcr
		\theguybelow\crcr}}}
\def\lta{\stacksymbols{<}{\sim}{2.5}{.2}}
\def\gta{\stacksymbols{>}{\sim}{3}{.5}}

\begin{document}

\title{CONFRONTATION OF INTRACLUSTER AND INTERSTELLAR GAS 
IN CLUSTER-CENTERED 
ELLIPTICAL GALAXIES: M87 IN VIRGO AND NGC 4874 IN COMA}

\author{Fabrizio Brighenti$^{1,2}$ \& William G. Mathews$^1$}

\affil{$^1$University of California Observatories/Lick Observatory,
Board of Studies in Astronomy and Astrophysics,
University of California, Santa Cruz, CA 95064\\
mathews@lick.ucsc.edu}

\affil{$^2$Dipartimento di Astronomia,
Universit\`a di Bologna,
via Ranzani 1,
Bologna 40127, Italy\\
brighenti@bo.astro.it}

%\vskip 2.in
%\noindent
%Received:

%\noindent
%PROOFS TO BE SENT TO:

%\noindent
%Lick Observatory

%\noindent
%Santa Cruz, CA 95064

%\noindent
%$^1$UCO/Lick Observatory Bulletin No.

%\vskip3.in
%\noindent
%Short Title: Elliptical Galaxies in Clusters
%\clearpage
\vskip .2in

\begin{abstract}

The evolution of hot interstellar gas in cluster-centered cD galaxies 
and the inflow of gas from the 
surrounding galaxy clusters are strongly coupled.
Cooling flows arise inside the cD galaxy 
because the deep stellar potential and stellar mass loss 
increase the gas density and decrease the radiative cooling time 
within the galaxy. 
Recent X-ray observations of M87 in the Virgo cluster 
and NGC 4874 in Coma reveal that the gas temperature 
beyond about 50 kpc from these cD 
galaxies is comparable to the virial temperature of the cluster, 
3 or 9 keV respectively, 
but within the optical galaxy the temperature drops to 
the galactic virial temperature $\sim 1$ keV. 
We show that these steep thermal gradients on galactic scales 
follow naturally from the usual cooling inflow 
assumptions without recourse to thermal conductivity. 
However, most of the gas must radiatively cool (``dropout'') 
before it flows to the 
galactic core, i.e. the gas must be multiphase.
The temperature and density profiles 
observed in M87 and NGC 4874 can 
be matched with approximate gas dynamical models calculated 
over several Gyrs with either globally 
uniform or centrally concentrated multiphase mass dropout.
Recent {\it XMM} observations of M87 indicate 
single phase flow at every radius with no 
apparent radiative cooling to low temperatures.
Gas dynamical models can be made consistent with 
single phase flow 
for $r \gta 10$ kpc, but to avoid huge central 
masses of cooled gas, we assume that some distributed 
cooling dropout occurs near the center of the flow where the 
gas temperature is $T \sim 1$ keV. 
The evidence in X-ray spectra for multiphase cooling 
beginning at lower temperatures $\sim 1$ keV
may be less apparent than for higher temperatures 
and may have escaped detection. 
However, even if the mass of cooled gas is distributed 
within $r \lta 10$ kpc, it is necessary that the mass of 
cooled gas not conflict with dynamical mass to light 
determinations.
Because of small deviations from true steady state 
flow, we find that 
the standard decomposition methods used by X-ray observers 
to determine the mass flow 
${\dot M}(r)$ may fail rather badly, 
particularly when the mass dropout decreases with 
radius in the flow.
For this case the decomposition procedure 
gives the usual cooling flow result, ${\dot M} \propto r$, 
which is quite unlike the true variation of 
${\dot M}(r)$ in our computed models.

\end{abstract}


\clearpage

%%%%%%%insert junktex.tex INTRODUCTION etc. here%%%%%%%%%%%

\section{INTRODUCTION}

Both galaxy clusters and individual
early-type galaxies contain virialized hot gas in which 
the radiative cooling time is much less than the Hubble time.
According to the standard interpretation,
as hot gas loses energy by emitting the X-radiation we observe,
it slowly migrates inward.
Since these ``cooling flows'' are nearly in hydrostatic
equilibrium in the cluster or galactic potential, 
the temperature and density profiles measured in the hot gas 
can be used to determine the underlying mass distribution. 
For this reason it is particularly interesting to 
study cooling flows in galaxies since we already 
have information about 
the mass distribution from optical studies within the 
half-light radius $r_e$ where the contribution of dark matter 
to the total mass is small.

Our objective in the calculations discussed here 
is to determine the nature of the 
interaction between cluster and interstellar gas 
in bright cD galaxies centrally located in galaxy clusters. 
In our previous studies of X-ray luminous massive elliptical 
galaxies located in small galaxy groups 
(e.g. Brighenti \& Mathews 1999), 
we have assumed that the interstellar and circumgalactic 
gas have been 
relatively unperturbed since the galaxy first formed. 
By following the growth of cosmological perturbations, 
including star formation and feedback, we can 
compute the gas flow over most of the Hubble time, 
successfully reproducing 
the radial dependence of hot gas density, temperature 
and abundance currently observed 
in massive group-centered elliptical galaxies.
However, the hot gas surrounding 
cluster-centered elliptical galaxies has a much higher 
cluster virial temperature.
As cluster gas flows into the central 
cD or elliptical galaxy, the interstellar gas 
is perturbed from that of 
more isolated or group-centered ellipticals, 
providing a useful test of the many parameters involved in these 
calculations. 
In particular, cluster-centered cD galaxies provide an excellent 
venue to explore the nature or reality of the mass ``dropout'' 
hypothesis, i.e. that the hot gas is
multiphase, having locally varying densities and cooling
times, so that it cools (``drops out'')
throughout a large volume of the cooling flow.
Unfortunately, the long term evolution of ambient cluster 
gas around cluster-centered cD galaxies is uncertain because of the 
likelihood of disruptive mergers in the cluster environment. 
Nevertheless, much can be learned if we assume that
the cluster gas as currently observed 
near the cD galaxy has been relatively stable for times
longer than several sound-crossing times, typically several Gyrs.

In the traditional interpretation of cooling flows,
it is assumed that the X-ray luminosity $L_x$ 
reflects energy lost from the hot gas 
which is not balanced by a heating source. 
As the radiating gas loses energy and 
slowly migrates toward the 
potential minimum, it is maintained near the local 
virial temperature by $Pdv$ compression. 
In the simplest (single phase) cooling flows 
all the gas cools rapidly at 
very center of the E or cD galaxy where the gas 
density and radiative emission are greatest.  
This model clearly fails since 
strong localized central peaks of thermal X-ray emission are not 
generally observed and the total
mass of cooled gas concentrated in the central regions would 
increase the stellar velocity dispersions 
considerably above those   
observed (e.g. Thomas 1986; 
White \& Sarazin 1987; Mathews \& Brighenti 2000).
To alleviate these difficulties, it has been fashionable 
for many years to assume that the hot gas 
is ``multiphase'' and therefore cools throughout 
a larger volume, depositing its mass in a less concentrated
manner (Nulsen 1986; Thomas et al. 1987; Allen et al. 2001).

Multiphase cooling can be incorporated into 
the equation of continuity for cooling flows 
with a mass sink term:
\begin{equation}
{\partial \rho \over \partial t} 
+ {1 \over r^2} {\partial \over \partial r}(r^2 \rho u) = 
\alpha_* \rho_* - q {\rho \over t_{cool}}. 
\end{equation}
The source term proportional to the stellar density $\rho_*$ 
represents stellar mass ejected from evolving stars.
The following dropout or 
sink term simply removes gas from the flow at a rate 
characterized by the local radiative cooling time 
at constant pressure 
$t_{cool} = 5 m_p k T/2 \mu \rho \Lambda$.  
This term is preceded by a dimensionless coefficient 
$q(r) \lta 1$ which is adjusted to best fit the 
observed X-ray surface brightness and other observations.
We recognize the {\it ad hoc} character of this sink term,
particularly in view of the well known failure of linear analysis to
find instability in cooling flows (e.g. Balbus \& Soker 1989).
Nevertheless, the compelling astrophysical arguments against
concentrated cooling at the center of the flow suggest either some
form of distributed mass dropout as in Equation (1) or an equally {\it
ad hoc} heating effect that supplies thermal energy without driving a
strong global wind.  Ironically, we have recently shown (Brighenti \&
Mathews 2002) that nonlinear density perturbations and subsequent
thermal instabilities occur naturally in cooling flows that are
decelerated by heating or additional 
non-thermal pressure near the center of the
flow.

To estimate $q$ we use the recent compilation 
of six well-observed cluster cooling flows 
from Allen et al. (2001) for which the local radiative 
cooling times are linear functions of radius, 
$t_{cool} = t_{cool,b} (r/r_b)$, for $r < r_b$.
Since $r_b \sim 40 - 120$ kpc is much greater than the 
central galactic half-light radius $r_e \sim 10$ kpc,
$\alpha_* \rho_* \ll \rho/t_{cool}$ beyond the central galaxy.
Multiplying the equation above by $4 \pi r^2 dr$, 
assuming quasi-steady state flow and substituting 
$\rho = 5 m_p k T/2 \mu t_{cool} \Lambda$, 
\begin{equation}
d {\dot M} = d(\rho u 4 \pi r^2)  
= - q {20 \pi m_p k T \over 2 \mu \Lambda}
{r^2 dr \over t_{cool}^2}.
\end{equation} 
We see that $t_{cool} \propto r$ implies 
${\dot M} \propto r$ as verified by Figure 6 of Allen et al. 
who determine 
consistent mass flow or dropout profiles ${\dot M}(r)$
from both spectroscopic and surface brightness observations.
Integrating this equation to $r = r_b$ and rearranging we find 
\begin{equation}
q = {2 \mu \Lambda(T) \over 20 \pi m_p k T} 
{{\dot M}(r_b) t_{cool,b}^2 \over r_b^3}
\end{equation}
which is constant in $r < r_b$ in this simple approximation.
Using data from the six well-observed clusters in Allen et al.,
we find $\langle q \rangle \approx 0.3 - 0.6$. 
But the implication that $t_{cool} \propto r$ requires 
${\dot M} \propto r$ is very sensitive to 
the steady state assumption as we show in Section 4 below.

Recently, {\it XMM} and {\it Chandra} have revealed new
and unexpected attributes of galactic and cluster cooling flows 
(McNamara et al. 2000; Fabian et al. 2001; Fabian 2001). 
In several cluster cooling flows there is spectroscopic 
evidence for only partial cooling, down to $\sim 2$ keV, 
but no further
(e. g. Peterson et al. 2001; Tamura et al. 2001; 
Kaastra et al. 2001; Molendi \& Pizzolato 2001;
Xu et al. 2001).
This has led some to question the reality of the 
$q$-term in the 
equation of continuity and to postulate instead 
some as yet poorly understood form of 
heating (e.g. Tucker \& Rosner 1983; Binney \& Tabor 1995;
Tucker \& David 1997; Ciotti \& Ostriker 2000; 
Soker et al. 2001). 
Although the traditional mass dropout hypothesis 
is being challenged, 
at the present time there is no accepted alternative 
model with heating 
that can be compared in detail with observations. 

Nevertheless, some recent observations continue to 
support the reality of traditional cooling flows. 
Buote (2000, 2001) finds evidence of widespread X-ray 
absorption from warm gas at $T \sim 10^6$ K in many 
cooling flows including M87.
FUSE observations of cluster A2597 
have detected the OVI doublet 
from gas at $T \sim 3 \times 10^5$ K 
corresponding to a lower than 
expected ${\dot M}$, but OVI lines are 
undetected in A1795 (Oegerle et al. 2001).
Similar 
FUSE observations of elliptical galaxies find OVI 
emission consistent with (but possibly 
somewhat less than) the expected 
mass cooling rate ${\dot M}$ and the usual multiphase
cooling flow model (e.g. Bregman et al. 2001). 
Perhaps some OVI line emission may have escaped detection 
because of broadening by turbulent velocities.  
Finally, Canizares et al. (1982) discovered line emission 
in M87 corresponding to temperatures of a few $10^6$ K 
which they interpret at radiatively cooling gas.
Although the evidence is somewhat contradictory at present,
we adopt provisionally the traditional assumption that 
cooling flows do in fact cool 
and seek values of $q(r)$ for cluster-centered cD galaxies 
that agree best with recent X-ray observations.

In the following we describe the transition flow as 
cluster gas enters M87 from the Virgo cluster 
and NGC 4874 from the Coma cluster, both of which have
been observed by {\it XMM} and/or {\it Chandra}. 
Remarkably, 
there is little evidence that the powerful, 
ongoing AGN in M87, with its jet and radio lobes, are 
heating the cooling flow at small galactic radii 
(e.g. Fabian 2001; Soker et al. 2001).
Indeed, we demonstrate below that the gas temperature, 
density and entropy profiles in M87/Virgo and 
NGC 4874/Coma are approximately 
consistent with a normal galactic-scale cooling flows, 
requiring little or no additional local heating. 
The interaction of cluster and galactic gas has been 
discussed previously 
by Thomas (1986), Meiksin (1990) and Bertin \& Toniazzo (1995).

\section{M87 IN VIRGO}

\subsection{M87/Virgo Mass Distribution}

Using ROSAT data, 
Nulsen \& B\"ohringer (1995) determined the 
azimuthally averaged mass distribution of the Virgo cluster 
within 50\arcmin~ (247 kpc at distance $d = 17$ Mpc) of M87.
Their best-fitting model has a total density structure 
\begin{equation}
\rho \propto [1 + (r/a)^2]^{-1}
\end{equation}
and mass profile 
\begin{equation}
M_{virgo}(r) = M_o[ r/a + \arctan (r/a) ]
\end{equation}
where $a = 47.6$ kpc and 
$M_o = 5.90 \times 10^{12}$ $M_{\odot}$.
The mass distribution is sensitive to the observed 
gas temperature variation $T(r)$ which 
for $r \gta 85$ kpc is approximately isothermal 
at $\sim 2.8$ keV ($3.2 \times 10^7$ K) 
(Nulsen \& B\"ohringer 1995).
The observed X-ray surface brightness 
in Virgo is more consistent with this $M_{virgo}(r)$ 
than with an NFW mass profile, although the difference 
is not great.

We determine the stellar mass distribution 
in M87 (NGC 4486) from the B-band surface brightness 
$\mu_B$ derived from the $\mu_V$ and B-V 
photometry of Goudfrooij et al. (1994) 
within $r = 132$\arcsec. 
Beyond this radius we use the V-band photometry 
of Schombert (1986) corrected to $\mu_B$ with B-V = 0.80, 
and brightened by $\delta V = 0.5$ to agree with the 
Goudfrooij et al. data in $r \lta 132$\arcsec.
The B-band surface brightness $\mu_B$ is transformed 
to surface mass density using $M/L_B = 8.2$  
(van der Marel 1991).
(Note that Faber et al. (1997) find a much larger 
value, $M/L_B = 23$.)
The stellar mass density, found by solving Abel's equation, 
can be fit with 
\begin{equation}
\rho_*(r) = (r/r_b)^{-1} [1 + (r/r_c)^{1.7}]^{-1}
[1 + (r/r_d)]^{-0.4}~~~M_{\odot}~{\rm pc}^{-3}
\end{equation}
where $r_b = 4.379$, $r_c = 1.513$ and $r_d = 15$ kpc.
The stellar mass distribution $M_{*,87}$ differs from 
a de Vaucouleurs profile because of the unusually 
large core (break) radius in M87 and an outer cD envelope 
of moderate brightness.
The potential is dominated by stars within 40 kpc. 
The density profile in Equation (6) is in good agreement 
with that of van der Marel (1994).
Finally, we assume that M87 is surrounded by 
a dark galactic halo with an NFW profile 
and total mass $M_{87,nfw} = 1.0 \times 10^{13}$ $M_{\odot }$. 
While it is unclear if it is necessary to append a separate 
dark halo to the central cD, we note that dynamical models 
of cluster-centered E galaxies have been unable to 
reproduce their massive cD envelopes (e.g. Dubinski 1998).
Fortunately, the mass $M_{87,nfw}$ has only a minor influence 
on our results because of the enormous mass of the 
Virgo cluster at every radius beyond the stellar 
half-light radius.

\subsection{Computational Procedure}

Each gas dynamical calculation begins with a determination 
of the gas density profile of isothermal intracluster 
gas in hydrostatic equilibrium in the 
Virgo potential defined by $M_{virgo}(r)$.
At cosmic time $t_i$ 
the central galaxy and its dark halo are inserted 
into the Virgo environment.
After this time stellar mass loss contributes
to the local gas at a rate
$\alpha_*(t) \rho_*(r)$ gm s$^{-1}$ cm$^{-3}$.
Here $\alpha_*(t) = \alpha_{*,n}
(t/t_n)^{-1.3}$ s$^{-1}$, appropriate for a 
nearly coeval burst of star formation at high redshift
where $\alpha_{*,n} = 4.7 \times 10^{-20}$ s$^{-1}$, and 
$t_n = 13$ Gyrs is the present age of the universe. 
Gas within the optical galaxy is heated by Type Ia 
supernovae which occur at a rate
SNu$(t) =$ SNu$(t_n)(t_n/t)$ where
SNu$(t_n) = 0.06$ supernovae per $10^{10}$ $L_B$ 
per 100 years. 
These stellar mass loss and supernova 
rates are consistent with 
X-ray observations of massive E galaxies
such as NGC 4472 that are not centered in large clusters 
(e.g. Brighenti \& Mathews 1999).

After time $t_i$, 
as the hot intracluster gas flows into the 
optical galaxy, it is cooled by 
radiation losses and mass ejected from evolving stars. 
Since mass-losing stars orbit rapidly 
through the hot gas, we expect instabilities to 
thermally mix stellar and ambient gases.  
This assumption is confirmed by negative abundance 
gradients of $\alpha$-elements observed in M87
(Molendi \& Gastaldello 2001) which cannot have
been produced by Type Ia supernovae. 
In large galaxies such as M87, the global potential 
is so deep that Type Ia heating has rather little 
effect on the thermal budget of the gas; 
the SNIa luminosity at $t_n$, 
$L_{SNIa} = 1.6 \times 10^{41}$
erg s$^{-1}$, is considerably less than the total 
X-ray emission $L_x$ from within the optical image of M87.
Following a brief transient ($< 1$ Gyr) just 
after M87 is introduced into the Virgo environment, 
the gas flow settles into a subsonic, but 
secularly changing flow. 
The gas dynamical equations that we use are identical 
to those discussed in Brighenti \& Mathews (1999) to 
which we refer the reader for further details. 
The equations are solved with a considerably modified 
1D spherical version of the ZEUS code (Stone \& Norman 1992) 
using 360 logarithmically increasing zones beginning 
with a central zone of size 15 pc and extending to 
2.1 Mpc. 
The appropriate initial temperature $T(t_i)$ for 
Virgo gas at time $t_i$ depends somewhat on the nature of 
the assumed mass dropout profile $q(r)$. 
If distributed mass dropout is small at large radii,
i.e. $q \approx 0$, the isothermal temperature 
$T(t_i)$ should be close to that observed by
Nulsen \& B\"ohringer.
But if $q \approx 1$ at large radii, the temperature 
of the background flow $T(t_i)$ must exceed that 
observed, which is a weighted average over both background 
and cooling regions. 

The factors that influence the radial 
variation of gas density and temperature within the 
optical galaxy at time $t_n = 13$ Gyrs are: 
(1) the assumed mass dropout profile $q(r)$ and 
therefore $T(t_i)$,
(2) the central gas density 
$n(r=0,t_i)$ of the hydrostatic cluster gas 
distribution just before the galaxy is inserted, 
(3) the time elapsed since this initial configuration  
$t_n - t_i$, 
(4) the specific stellar mass-loss rate $\alpha_{*}$ and 
(5) thermal conductivity.
As we show below, our computed values of
$n(r,t_n)$ and $T(r,t_n)$ inside and near M87 
are only weakly sensitive to 
the time $t_n - t_i$ over which the gas flow evolves 
after the initial setup, provided $t_n - t_i$ is larger 
than $\sim 3$ Gyr, i.e. several sound crossing times. 
The time $t_n - t_i$ can be regarded as the (unknown) interval 
since the last major merger event in Virgo when the 
gas density and temperature may have changed appreciably.
While it is unlikely that $\alpha_*$ varies strongly with 
galactic radius, 
the normalization coefficient $\alpha_{*,n}$ can influence 
$n(r,t_n)$ and $T(r,t_n)$ somewhat. 
It is unlikely that thermal conduction plays an important 
role in the thermal structure of
M87/Virgo (e.g. Saito \& Shigeyama 1999). 
We find that thermal conduction at $\gta 0.1$ of the 
Spitzer (or saturated) rate cannot occur since the 
gas temperature would be nearly isothermal 
at $T \approx T(t_i) \sim 3$ keV 
even within $\sim 30$ kpc and this is not observed. 
Thermal conduction can help reduce the total 
mass of cooled gas in the central regions of M87, 
but this requires that the conductivity be reduced by 
a carefully chosen factor 
that preserves the observed temperature profile 
(e.g. Bertschinger \& Meiksin 1986). 
Therefore, in our discussion below we concentrate mostly 
on the remaining factor that influences 
the currently observed gas density and temperature 
distributions: the mass dropout coefficient $q(r)$.

\subsection{Results}

In Figures 1 and 2 we compare the observed gas density and 
temperature profiles in M87/Virgo with 
results of our gas dynamical calculations. 
Unless otherwise noted,
all calculations begin at cosmic time $t_i = 6$ Gyrs and 
continue to the present time $t_n = 13$ Gyrs.
At time $t_i$ the hot cluster gas is 
in hydrostatic equilibrium in the mass distribution 
$M_{virgo}(r)$.
The cluster gas is 
isothermal at $T(t_i) = 3$ keV ($3.5 \times 10^7$ K) 
for models with $q(r) \ll 1$ at $r \gta 20$ kpc and 
$T(t_i) = 3.4$ keV ($4.0 \times 10^7$ K)
for (globally multiphase) models with $q = 1$.
The central gas density $n(0,t_i) = 0.003$ cm$^{-3}$ is 
chosen to agree with X-ray surface brightness 
observations at radii far beyond the optical extent of M87.

If $q = 1$ throughout the flow, about half of the
total X-radiation comes from the cooling elements and 
the rest comes from the background flow.
We distinguish these two sources of emission 
in Figures 1 and 2
by plotting both the background density 
$n_h(r)$ (light solid line) and the density $n(r)$ 
that would be determined from the combined emission from 
background and cooling regions (heavy solid line).
This latter $n(r)$ can be compared with the density profile 
determined from observations.
In the temperature plots we show three temperatures:
the temperature 
$T_h(r)$ of the hot background gas flow at radius $r$
(dotted lines), 
the mean local emission-weighted temperature $T(r)$ 
including both background and cooling gas 
(light solid lines), and the 
emission-weighted temperature $T(R)$ which is $T(r)$ 
averaged along the line of sight, including emission from 
both background flow and cooling gas (heavy solid lines).
$R$ is the projected radius. 
If $dT/dr > 0$, $T(R)$ is greater than $T(r)$ 
for $r = R$ because 
emission from hotter gas along the line of sight is included 
in the weighted average $T(R)$.
The temperature $T(R)$ reported by observers is 
not simply an emission-weighted mean along the line of sight,
but is derived by fitting single temperature (1T) or 
composite temperature (2T or 1T + CF) XSPEC models to 
the projected X-ray spectrum. 
Nevertheless, in the following discussion we assume that our 
computed $T(R)$ -- which includes effects of $dT/dr \ne 0$ 
as well as cooling dropout when $q \ne 0$ -- 
can be directly compared with observed temperatures.

Over time $t_n - t_i = 13 - 6 = 7$ Gyrs,
cooling flows do not form in Virgo without M87. 
If the evolution of the cluster gas after $t_i$ is calculated 
without M87, we find that $T(R)$ at time $t_n$
rises from $\sim 1.5$ keV at the center, reaches 
2.6 keV at $r = 100$ kpc, then slowly rises toward $T(t_i)$. 
The gas density is similar to $n(r,t_i)$ for $r \gta 50$ kpc,
but within this radius it is much less than observed,
$n(r,t_n) \lta 0.01$ cm$^{-3}$ 
and by time $t_n$ no gas has cooled.
However, when M87 is introduced at $t_i$, the central gas density 
and X-ray surface brightness 
increase both because of the locally deepened 
(stellar) potential 
and because of the addition of stellar mass loss.
We find that the deep potential of M87 
is sufficient to create a cooling flow in M87/Virgo 
even if stellar mass loss is absent; this 
is possible since the Virgo virial temperature is 
only 2 - 3 times larger than that of M87.
Stellar mass loss further cools the 
cluster gas and increases its density 
as it flows into the optical galaxy,
lowering the radiative cooling time.

\subsubsection{Gas Flow with Constant and Variable $q$}

Figure 1 shows the resulting density and temperature 
profiles for the $q = 1$ model in which 
cooling dropout occurs uniformly throughout the flow.  
The computed density variation $n(r)$ 
(heavy solid line) agrees quite well with 
the gas density observed in M87/Virgo. 
The density continues to rise in $r < 1$ kpc, 
where the observational data are sparse,
but in this region it is likely that the magnetic 
field is high enough $\sim 100\mu$G to flatten the gas 
density profile. 
Owen, Eilek \& Keel (1990) estimate 
a field of $\sim 40\mu$G at $r = 2$ kpc. 
The projected model temperature profile $T(R)$ 
also agrees reasonably well with observed values 
but is somewhat lower 
for $r \lta 2$ kpc and $r \gta 45$ kpc. 
M87 observations are more difficult near the center 
because of non-thermal X-radiation from the 
active nucleus (e.g. B\"ohringer et al. 2001;
Molendi \& Gastaldello 2001),  
which may explain some of the 
temperature discrepancy at small $r$, 
along with a variety of other possible explanations (see below).
The failure of the $q = 1$ model to 
match observed temperatures beyond $\sim 45$ kpc suggests that 
$q$ may be less than 1 in this region since 
the observed temperatures in Figure 1  
increase toward that of the uncooled background gas
$T_h(r)$ (dotted line) which is $\sim 0.8$ keV hotter than $T(R)$.

To explore variable mass dropout, we show in Figure 2 the results 
of a calculation with a cooling dropout 
that decreases with gas density and 
therefore galactic radius, $q = {\rm exp}[-0.05/n(r)]$, 
but which is otherwise identical to the solution
shown in Figure 1.
As a function of radius, $q \approx 1$, 0.5 and 0.1 at 
$r = 0.0$, 2.8 and 7.6 kpc respectively. 
The agreement with observed density is very good and
$T(R)$ agrees adequately with observed values.
It is interesting that the observed density $n(r)$ and 
temperature $T(R)$ profiles can be 
fit for either constant or spatially 
variable $q$, particularly for $r \gta 10$ kpc.
But the observed spectra 
of the two flows are quite different, as we discuss below.

Inside $\sim 2$ kpc in Figure 1 or $\sim 10$ kpc in Figure 2
the computed temperature $T(R)$ is somewhat lower than observed.
It is unlikely that this discrepancy is due to shock 
heating by the M87 AGN or radio source. 
In our experience the principal 
long term consequence of local heating 
is to lower the central gas density with only a very small 
change in temperature.
In any case, {\it Chandra} observations provide 
little or no direct evidence of shock heating in radio 
galaxies (e.g. Fabian 2000). 
Several possible explanations for this temperature 
discrepancy are explored in the additional 
$q = {\rm exp}[-0.05/n(r)]$ solutions shown in Figure 3.
If the time since the initial M87/Virgo configuration 
$t_n - t_i$ is $\lta 4$ Gyrs, we find that both $T(R)$ and 
$dT/dR$ are larger in $R \lta 10$ kpc. 
This is illustrated in Figure 3 with a solution 
at $t_n - t_i = 13 - 10 = 3$ Gyrs (long dashed lines)
which is a poor fit to observed $n(r)$ and $T(r)$.
However, for $t_n - t_i \gta 4$ Gyrs 
all $T(R)$ profiles resemble that in Figure 2.
Reducing the normalization of the 
stellar mass loss rate $\alpha_{*,n}$ by 
a factor of 2 also has a beneficial, 
but limited affect on $T(R)$ (short dashed line in Fig. 3).
However, if the stellar mass to light ratio is 
twice that of van der Marel, $M/L_B = 16.4$ 
-- and therefore in better agreement with the $M/L_B$ of 
Faber et al. (1997) --  
the temperature profile is considerably improved 
(solid line).
(Note: $\alpha_* \propto (M/L_B)^{-1}$ 
has been reduced by 2 in this model.) 

In Figure 2 $T(R)$ is in excellent agreement with observations 
in $R \gta 10$ kpc.
In this region the gas is essentially single phase, 
$T_h(r) = T(r)$, and the full line of sight variation  
$\sim 2[T(R) - T(r)] \approx 0.4 - 0.7$ keV is comparable to 
the cooling flow temperature range 
$T_{max} - T_{min}$ in the XSPEC (Model C) analysis 
of Molendi \& Pizzolato (2001) at $r \gta 10$ kpc. 
The multiphase appearance is due entirely to contributions 
of various single phase temperatures along the line of sight. 
At smaller radii, $r \lta 10$ kpc, $q$ increases, 
and $T(R)$ is reduced by contributions from gas that 
is cooling below 1 keV. 

We have also repeated M87/Virgo calculations with no dropout 
at all, $q = 0$, so that all the gas cools right at the origin. 
The results of these calculations agree surprisingly well 
with observations at $r \gta 1$ kpc, 
particularly if $\alpha_*$ is also lower.
But there is a major problem.
In $r \lta 1$ kpc both the gas density and temperature 
increase in the (point source) potential of the 
centrally cooled mass, producing a dense, luminous X-ray core.
However, central masses $\gta 10^{10}$ $M_{\odot}$ 
and central peaks in $T$ and $L_x$ 
are not typically observed, so cooling inflows without 
spatially distributed dropout are unrealistic. 

\subsubsection{Multiphase flow in M87/Virgo}

The detailed shape of the density and temperature profiles 
in M87/Virgo 
can be adequately reproduced with either constant 
or variable $q$, 
but this degeneracy can be broken by examining the
X-ray spectra of these models in more detail.
If $q = 1$ as in Figure 1, 
the line and continuum X-ray spectrum in $r \gta 20$ kpc 
should consist of two components 
of approximately equal flux:
(1) a blend of emission from line of sight plasma 
at temperatures $T_h \sim 2.8 - 3.4$ keV, and 
(2) emission from regions cooling at every radius 
through all temperatures $T < T_h$. 
However, if $q \approx 0$ as in our variable $q$ model 
beyond 10 kpc,
$T(R)$ is determined by only the first of these two 
components.
In the upper panel of Figure 4 we show 
approximate local spectra at $r = 30$ kpc for both  
constant and variable $q$ models.
For the $q =  {\rm exp}[-0.05/n(r)]$ flow 
we use a single temperature (1T), solar abundance 
mekal model 
(using XSPEC version 11.0.1) at $T = 2.5$ keV. 
For the $q = 1$ model spectrum 
we combine two bolometrically equal components: 
a 3 keV 1T spectrum and 
a mekal cooling flow (CF) spectrum with $T_{max} = 3$ keV.
The difference between the 
$q = 1$ and variable $q$ spectra is most apparent near 
the Fe-L line complex near 1 keV 
which is broadened toward lower 
energies in the (1T + CF) spectrum for $q = 1$ flow.

The current observational 
evidence for truly multiphase gas in M87/Virgo
is contradictory. 
Buote (2001) reanalyzed {\it ROSAT} PSPC data from 
M87/Virgo, 
including the important spectral range down to 
0.2 keV, and concluded that a single temperature gas 
cannot explain the X-ray spectrum.
Instead, Buote finds evidence of emission 
in the 0.2 - 0.4 keV band from warm 
($T \sim 10^6$ K) gas in M87/Virgo at almost every 
radius $\lta 70$ kpc except, curiously, in the 
$r = 5 - 10$ kpc region where the multiphase 
flow appears to be suppressed. 
Buote also finds 
(oxygen edge) absorption in the 0.5 - 0.8 keV range 
throughout M87/Virgo consistent with
absorption by the same warm gas that emits at lower energies.
The mass of warm gas necessary for this absorption 
in M87 is enormous,
comparable to the total mass of hot gas observed, 
and it is unclear why the warm gas would not cool further
(see also Bonamente et al. 2001).
An important part of Buote's argument, shown in 
his Figure 4, is that if data below 0.5 - 0.7 keV is 
neglected, as many authors have done,
the absorbing foreground column
$N_H$ greatly exceeds the Galactic value;
conversely, if the entire spectrum down to 0.2 keV 
is considered, the foreground $N_H$ is approximately Galactic 
at every projected radius in M87/Virgo, 
but some warm gas absorption is required. 
Buote's {\it ROSAT} data cannot easily discriminate 
between spectra of two temperature plasma 
and and a single temperature with cooling elements.

This evidence for multiphase gas in M87/Virgo 
is not supported by {\it XMM} data analyzed 
by B\"ohringer et al. (2001) 
which are spectroscopically superior to {\it ROSAT}.
When they attempted a multiphase solution by 
combining spectra from $T_h$ with that of a cooling gas,
B\"ohringer et al. found mass deposition rates 
of ${\dot M} \sim 10 \pm 5$ $M_{\odot}$ yr$^{-1}$
within 50 kpc and absorbing columns 
$N_H \sim 5 \times 10^{21}$ cm$^{-2}$ that are 
about 30 times the Galactic value.

Molendi \& Pizzolato (2001) have reanalyzed 
0.5 - 4.0 keV {\it XMM} EPIC data from M87/Virgo and 
lend support to the conclusions of B\"ohringer et al.:
(1) the only multiphase attributes apparent in  
M87/Virgo spectra are those introduced by 
the variation of single phase temperatures along the 
line of sight, and 
(2) there is no evidence for 
intrinsic multiphase gas that is cooling locally.
In their best multiphase cooling flow (Model B), 
Molendi \& Pizzolato find 
(1) that the (spectroscopic) 
total mass deposition rate in M87/Virgo is 
less than 0.5 $M_{\odot}$ yr$^{-1}$, at least an order of 
magnitude less than that expected from $L_x$, and
(2) very large absorbing columns $N_H$. 
They suggest that earlier claims of large soft X-ray 
absorption
are an artifact resulting from the multiphase assumption. 
But there remain problems with the model favored
by Molendi \& Pizzolato, 
Model C, in which the temperature range 
viewed along the sight
is sufficient to account for the full temperature 
range observed at all $R$.
They model this temperature 
range with a partial isobaric cooling flow 
from $T_{max}$ to $T_{min}$ rather than by integrating 
over the known $T(r)$ along the line of sight. 
But gas cannot just cool to some minimum temperature 
-- ether locally as cooling dropout 
or during its radial flow -- 
and accumulate there since that would eventually 
distort the entire spectrum toward this low temperature.
This is the conundrum discussed by Fabian et al. (2001)
who claim that ``cooling flow'' gas appears to cool to 
some intermediate temperature $T_{min}$, then disappear.

However, we must not overlook the 
{\it Einstein} FPCS observation 
of Fe XVII - Fe XXIV lines in the central $\sim 2$\arcmin 
($\sim 10$ kpc) of M87 by 
Canizares et al. (1982) that reveal gas temperatures 
ranging from $3 \times 10^7$ K down to $2 - 4 \times 10^6$ K. 
The volume emission measure from cooling gas suggests 
a cooling rate of 3 - 4 $M_{\odot}$ yr$^{-1}$.
This is clear evidence in support of the 
multiphase model, at least in this central region.

\subsubsection{Spectroscopic Evidence for Variable $q$}

Returning to Figure 4, the easily visible low-energy extension 
of the Fe-L line feature, which indicates cooling to very 
low temperatures, was evidently not seen at $r \sim 30$ kpc  
by Molendi \& Pizzolato (2001) in their XMM spectra of M87.
This is in agreement with our variable $q(r)$ model  
in which $q \approx 0$ in $r \gta 10$ kpc. 
However, 
in their (preferred) Model C analysis of {\it XMM} data 
in which the temperature cools through a finite range, 
simulating the variation of $T$ along the line of sight, 
Molendi \& Pizzolato find temperature ranges of
$1.0 < T < 1.8$ keV at $r = 3.9$ kpc and 
$1.4 < T < 1.9$ at $r = 7.5$ kpc. 
These narrow ranges are inconsistent with intrinsic 
multiphase cooling in this region,  
with the observations of Canizares et al. 1982,
and with our variable $q$ model.
However, it may be more difficult to detect 
multiphase cooling when the gas cools 
from low temperatures, $T \sim 1$ keV, as in our 
variable $q$ model within 10 kpc.

To illustrate this, in the lower panel of Figure 4 
we use XSPEC (ver. 11.0.1) to compare  
intrinsic 0.2 - 3 keV spectra at $\sim 1$ kpc, 
assuming solar abundances. 
The solid line shows the combined 
spectrum of a single temperature (1T) gas at $T_h = 0.95$ keV
plus a cooling flow (CF) 
spectrum for $T = 0.95 \rightarrow 0.1$ keV, 
with equal normalizations.
For comparison we show with a dotted line 
in Figure 4 the mekal 
spectrum of a single phase gas at the mean temperature 
of the 1T + CF flow, $T \approx 0.8$ keV (cf. Fig. 2).
The difference between these two spectra 
is much less than that between 1T and (1T$_h$ + CF) 
spectra at $\sim 3$ keV (upper panel) 
corresponding to $R = 30$ kpc
where the Fe-L feature is widened by the cooling gas. 
To demonstrate conclusively 
that gas at $\sim 1$ keV
is not cooling to very low temperatures,
it would be necessary to be confident of the relative 
strengths of the Fe-L lines in 0.7 - 1.0 keV which 
may not be possible at present.
Alternatively, 
the line ratio OVII(0.574 keV)/OVIII(0.652 keV) increases  
with decreasing temperature below 1 keV
(Landini \& Monsignori Fossi 1990), 
which is apparent in Figure 4. 
However, OVII(0.574 keV)/OVIII(0.652 keV) could 
be artificially increased 
by oxygen-edge absorption (Boute 2001), 
which is a further possible complication.
On balance, therefore, these spectroscopic 
considerations favor the variable $q(r)$
model for M87/Virgo,
in which $q$ becomes effectively nonzero only
within 10 kpc where a multiphase spectrum would be expected, 
but which may be difficult to observe.

\subsubsection{Where is the Cooled Gas?}

A potentially serious 
problem with the variable $q$ model is that the 
centrally concentrated mass dropout may conflict with 
the stellar mass to light ratio of M87.  
In his global dynamic models for M87, 
van der Marel (1991) found $M/L_B = 8.2$ 
while Faber et al. (1997) find $M/L_B = 23$ using 
the King method applied to the galactic core
(Richstone \& Tremaine 1986; Tremaine et al. 1994).
In these estimates 
$M/L_B$ is always assumed to be uniform with 
radius -- which would not be expected if 
mass dropout is important. 
The discrepancy in $M/L_B$ 
could indicate that dark dropout 
mass exists mostly within the M87 core, 
although van der Marel (1994) also finds  
a stellar $M/L_B = 8.2$ in stars near the 
central black hole and 
Merritt \& Oh (1997) find $M/L_B = 11$. 
The black hole mass considered by van der Marel, 
$M_{bh} = 3 \times 10^9$, is only three times less 
than the total stellar mass 
$M_b = 1.0 \times 10^{10}$ $M_{\odot}$ within the core 
(or break) radius $r_b = 626$ pc (7.61\arcsec).
This black hole mass is confirmed by observations
of a thin HII disk in Keplerian rotation to 
within 4 pc of the center 
(Marconi et al. 1997; Macchetto et al. 1997).

Most of the dark matter deposited in the M87 core 
by the variable $q$ solution occurs at very early times.
To study this in more detail 
we use an ``alternative'' M87/Virgo model in which 
the initial evolution of M87 occurs outside
Virgo and the galactic scale 
cooling flow deposits mass according to 
$q = {\rm exp}(-0.05/n)$. 
By time $t_i = 6$ Gyrs when M87 is inserted in Virgo, 
the total mass of cooled (dropped out) gas is 
already $M_{do} = 2.3 \times 10^{10}$ $M_{\odot}$
or $2.3 M_b$.
At these early times our schematic model for galaxy 
evolution is very uncertain. 
Since the stars in M87 have significant metal enrichment, 
some of the enriched galactic hot 
gas that cooled at early times must have 
formed into second 
generations of old luminous stars observed today
(Nulsen \& Fabian 1997). 
But it is unclear how long cooled gas formed 
into optically luminous stars.
Many authors 
have suggested that at the present 
time only low mass, optically dark stars are forming 
in cooling flows, 
but other lines of evidence, such as the stellar 
H$\beta$ index, suggest that luminous stars continue 
to form at the present time (Mathews \& Brighenti 1999).
Furthermore, the precise value of the total 
mass of cooled gas at time $t_i$ depends critically 
on the unknown cooling and merger history of M87 
before $t_i$ 
and whether or not $q = {\rm exp}(-0.05/n)$ applies at 
these early times, as we have assumed.
By time $t_n = 13$ Gyrs, in the variable $q$ flow 
$5.3 \times 10^{10}$ $M_{\odot}$ has cooled.

If the cooling dropout hypothesis is correct, 
the currently observed 
stellar mass to light ratio must reflect in part 
a contribution from cooled gas. 
If luminous stars form continuously from cooled 
gas, the mass to light ratio could in principle 
be similar to that of the old stellar population, 
although an exact match seems unlikely. 
On the other hand, if only non-luminous stars formed 
in our $q = {\rm exp}(-0.05/n)$ flow, 
then at time $t_i = 6$ Gyrs 
$M/L_B$ would have already increased to 18 
within the break radius
$r_b$, ignoring passive cosmological luminosity 
evolution. 
By time $t_n = 13$ Gyrs, $M/L_B$ within 
$r_b$ increases
further to 28 which is unacceptably high.
Notice, however, that in our models 
the mass dropout continues to increase
with rising gas density in $r \lta r_b$ 
where about half of the dropout occurs 
for the variable $q$ flow.
It is more likely that the gas density in M87 levels
off in $r \lta 1$ kpc similar to other large 
elliptical galaxies in Virgo
(e.g. Brighenti \& Mathews 1997), possibly due to 
large central magnetic fields or the central AGN. 
If so, mass dropout within $r_b$ would be much reduced. 
More generally, the success of the variable 
$q$ flow does not depend critically on the exact profile 
of $q(r)$ or $M_{do}(r)$ within $\sim 7$ kpc. 
To consider the most optimistic case,
suppose all the cooled gas at time $t_n$
were distributed uniformly with old stars, 
$M_{do}(r) \propto M_*(r)$, 
within 7 kpc, then the central $M/L_B$ 
would increase from 8.2 to only 8.6. 
In order to accommodate any increase in $M/L_B$, 
it is necessary to assume  
that the underlying old stars in 
M87 have $M/L_B$ less than our adopted 8.2 and  
that the dark dropout has increased it to this value. 
Finally, mass dropout in the $q = 1$ flow is more 
spatially diffuse: for optically dark dropout 
the mass to light ratio within $r_b$ is
$M/L_B = 12$ at $t_i$ and
$M/L_B = 14$ at $t_n = 13$ Gyrs.

It is well known that thermal conductivity can reduce 
the cooling rate in the centers of central cD galaxies 
in galaxy clusters, but the 
thermal conductivity must be fine-tuned 
to lower ${\dot M}$ without 
unrealistically flattening the temperature gradient 
(Bertschinger \& Meiksin 1986; Meiksin 1988; 
Bregman \& David 1988; Rosner \& Tucker 1989). 
We have repeated $q = 1$ and variable $q(r)$ 
calculations for M87/Virgo, assuming the ``alternative'' 
evolution discussed above and including thermal conduction.
We assume that the Spitzer or saturated 
thermal conductivity is reduced uniformly by a factor $f$ 
to simulate suppression by tangled magnetic fields.
In Figure 5 we show that the density and temperature 
profiles in M87/Virgo can be fit reasonably well 
with $f = 0.02$ for $q = 1$ flow and 
$f = 0.08$ for variable $q$ flow.
If $f$ is less than these values, the conductivity has 
little effect and the final flows resemble those described 
earlier without conduction.  

Conductive heating reduces cooling dropout.
For the variable $q$ solution with $f = 0.08$ 
the total mass of cooled gas is $3.1 \times 10^{10}$ 
$M_{\odot}$, about half that without conductivity. 
Unfortunately, these results are very sensitive to 
the factor $f$. 
The dashed lines in 
Figure 5 show that if $f$ is changed to 0.05 ($q = 1$) 
or 0.15 (variable $q$), the gas temperature is 
unacceptably high near the center. 
Since there is no obvious reason why the 
conductivity should be lower by just the factor $f$
required to reduce $|{\dot M}|$, these models
cannot be accepted.

%\end{document}

\section{NGC 4874 in Coma}

Evidence that the Coma cluster is not in dynamical equilibrium 
was noticed at {\it ROSAT} resolution by 
Briel, Henry \& B\"ohringer (1992) and Briel \& Henry (1997)
who found significant subclustering and asymmetric 
temperature variations. 
Direct {\it ROSAT} images (Dow \& White 1995) 
and the wavelet analysis of X-ray data by 
Biviano et al. (1996) clearly show that 
the center of Coma consists 
of two subclusters, each centered on the bright 
cD elliptical galaxies NGC 4489 and NGC 4874, 
separated in projection by 7.1\arcmin~.
Recent {\it XMM} data confirm the existence 
of enhanced, $\sim 100$ kpc 
X-ray emission around these two galaxies 
(Arnaud et al. 2001; Briel et al. 2001) 
and so do the {\it Chandra} images of 
Vikhlinin et al. (2001). 
We assume here that the two bright E galaxies are 
each centers of merging subclusters 
so the ambient dark matter and gas peaked around 
each galaxy is locally comoving with the 
galaxies as they orbit in Coma, i.e. there is no 
local ram pressure effect near the cD galaxies. 
The transient, dynamically disturbed 
nature of Coma is consistent 
with the observation that Coma is among the minority of 
clusters that do not have the characteristic 
centrally peaked X-ray cooling flow structure. 
In view of the unrelaxed nature of Coma, we begin 
our gas dynamical calculations at a more recent 
cosmic time $t_i = 10$ Gyrs, but $t_n - t_i$ is 
still longer than the relevant sound crossing times.

The gas temperature 
within a few kpc of NGC 4489 and NGC 4874 
decreases dramatically 
from the virial temperature of the Coma cluster,
$T \approx 8 - 9$ keV at $r \gta 7$ kpc 
to the virial temperature of the 
embedded galaxy $T \sim 1$ keV at $r \lta 4$ kpc.
(Arnaud et al. 2001; Vikhlinin et al. 2001).
Because of its larger temperature excursion, 
Coma is a more extreme example 
of the interaction between cluster and galactic 
gas than M87/Virgo.
Yet, as we show below, 
the steep temperature gradient observed 
can also be understood 
with the classical cooling flow hypothesis. 
According to the {\it Chandra} observations of 
Vikhlinin et al (2001), the emission-weighted 
temperature near NGC 4874 should increase from 
\begin{equation}
\langle T_1 \rangle = 1.0 \pm 0.04~ {\rm keV}~~~
{\rm inside}~~~ 7\arcsec~ (3.40~ {\rm kpc}) 
\end{equation}
to
\begin{equation}
\langle T_2 \rangle = 8.9 \pm 0.4~{\rm keV}~~~
{\rm in}~~~10\arcsec - 60\arcsec~ (4.86 - 29.2~ {\rm kpc}).
\end{equation}
\vskip.1cm
$\langle T_2 \rangle$ is a line of sight average.
In deriving $\langle T_1 \rangle$, 
Vikhlinin et al. removed background emission from 
15 - 30\arcsec, but not the local background 
in the 7 - 15\arcsec region.
Nevertheless, $\langle T_1 \rangle$ approximately 
represents an average temperature over the physical 
radius interior to 7\arcsec~ or 3.4 kpc.
We assume a distance to Coma of 106 Mpc 
based on $H_o = 70$
km s$^{-1}$ per Mpc and v = 7461 km s$^{-1}$ 
(Faber et al. 1997). 
At this distance 1\arcsec = 0.486 kpc.

\subsection{NGC 4874/Coma Mass Distribution}

The two bright elliptical galaxies in Coma have very similar 
optical and X-ray properties. 
However, we concentrate here on NGC 4874 because it 
is more centrally located (in projection) in the cluster 
and the surrounding X-ray isophotes are more symmetrical.
We assume that the mass distribution is also spherically 
symmetric about NGC 4874. 
Since our interest is in the gas flow pattern 
within $\sim 100$ kpc from the center of NGC 4874, 
it is not necessary to approximate  
the complete mass distribution of Coma 
with high precision.
We adopt a large scale isothermal total mass profile
$M_{coma}(r)$ as given in Equation (5) with  
$M_o = 2.8 \times 10^{14}$ $M_{\odot}$ and core 
radius $a = 300$ kpc, which is atypically large 
due to dynamically transient substructure.
This mass confines the cluster gas with an X-ray 
surface brightness distribution similar to that 
observed by Briel et al. (1992) out to $r \sim 1$ 
Mpc and beyond. 
In addition, we adopt an NFW dark matter halo for the 
galaxy, $M_{4874,nfw} = 1 \times 10^{13}$ $M_{\odot}$, 
although this mass has almost no effect on our solutions 
since the galactic virial temperature is so much less 
than that of the Coma cluster.

The stellar mass distribution 
is found from the R-band surface brightness measurements 
of Peletier et al. (1990), which vary linearly with the 
log of the projected radius $R$, 
\begin{equation}
\mu_R = 17 + 3.125 \log R~~~~~~{\rm mag/}\Box{\rm \arcsec}. 
\end{equation}
The corresponding B-band brightness $\mu_B$ is found  
assuming $B - R = 1.67$ typical for E galaxies 
(Fukugita, Shimasaku \& Ichikawa 1995).
As before we convert $\mu_B$ into physical units and 
invert with Abel's equation to find the following 
approximate stellar space density
\begin{eqnarray}
\rho_*(r) &  = & \rho_*(r_b) (r/r_b)^{-0.76}  ~~~r < r_b\\
          &  = & \rho_* (r_b)
( {r / r_b})^{-2.27} ~~~ r_b <  r < r_{max}
\end{eqnarray}
where $\rho_*(r_b) = 2.137$ $M_{\odot}$ pc$^{-3}$, 
$r_b = 2.67$\arcsec~ (1.30 kpc) is the break radius 
(Faber et al. 1997) and $r_{max} = 33.71 r_b$. 
This $\rho_*(r)$ is very close to the stellar mass density 
shown in Figure 4 of Harris et al (2000) and we have 
used $M/L_B = 8$ from their paper.
The stellar mass distribution of NGC 4874 dominates
the gravitational potential in $r \lta 12$ kpc.

\subsection{Results}

As before, we begin our NGC4874/Coma calculation 
by establishing an isothermal 
gas background in the cluster 
with temperature $T_{coma} \approx 8 - 9$ keV.
Initially the gas is in hydrostatic equilibrium 
in the approximate dark halo of Coma, $M_{coma}(r)$. 
At time $t_i = 10$ Gyrs we place NGC 4874 and its dark 
halo at the center of the cluster, and begin the 
gas dynamical calculation with the same stellar 
mass loss and supernova parameters that we used for 
M87/Virgo. 
We adopt a more recent starting time $t_i = 10$ Gyrs 
for NGC4874/Coma because Coma is dynamically unrelaxed
compared to M87/Virgo.
The temperature $T_{coma}$ and central density 
$n(0,t_i)$ at time $t_i$ determine the pressure of the 
cluster gas near NGC 4874 
and are adjusted to agree with the large scale 
gas density distribution in Coma.

For a fully multiphase flow model with $q = 1$, 
we use $n(0,t_i) = 0.0015$ cm$^{-3}$ and 
$T_{coma} = 13$ keV ($1.5 \times 10^8$ K).
When $q = 1$, the background 
cluster temperature must be significantly higher 
than temperatures observed since the latter also 
includes cooling gas at every radius. 
Figure 6 shows the gas electron density and temperature 
for the $q = 1$ model after it has evolved to 
time $t_n = 13$ Gyrs.
In discussing their {\it Chandra} observations of NGC 4874, 
Vikhlinin et al. (2001)
show the X-ray surface brightness distribution $\Sigma_x$, 
but not the density profile $n(r)$.
Since NGC 4874 was 2.8\arcmin~ from the 
center of the {\it Chandra} field, 
the spatial resolution is degraded 
to a half-energy radius of $\sim 0.9$\arcsec.
Figure 7 shows how the X-ray surface brightness   
$\Sigma_x$ (0.2 - 5 keV) 
of our $q = 1$ solution (solid line) is altered 
by such a Gaussian point spread function (dashed line)
using the integration procedure of Pritchet \& Kline (1981).
In Figure 7 we plot the (unnormalized) 
surface brightness observations of Vikhlinin et al. (2001) 
(circles) and Dow \& White (1995) (squares) by adjusting 
the data vertically for the best fit.
The fit is acceptable except perhaps near $R \sim 2 - 3$ kpc
where the model emissivity is too high, but this is the 
region where {\it Chandra} observations become less accurate
in 0.5 - 2 keV.

The physical $T(r)$ 
and projected $T(R)$ temperature profiles 
for the $q = 1$ calculation shown in Figure 6 
can be used to determine 
the mean temperatures in the two regions 
in Equations (7) and (8):
$\langle T_1 \rangle = 1.0$ keV and 
$\langle T_2 \rangle = 7.8$ keV.
The latter temperature is less than 
the value  $\langle T_2 \rangle \approx 8.9$ keV 
observed by Vikhlinin et al. (2001), but is in better 
agreement with temperatures from Arnaud et al. (2001) 
in this same region.
$\langle T_2 \rangle$ is also much less than the assumed 
temperature in the background gas, $T_{coma} = 13$ keV. 
Also shown as filled circles in Figure 6 are local 
temperatures determined by Vikhlinin et al. which agree 
rather well with the (unprojected) $T(r)$ in this model.

We also consider an identical flow, 
but with spatially variable dropout,
$q = {\rm exp}[-0.05/n(r)]$, and initial density 
$n(0,t_i) = 0.003$ cm$^{-3}$ and 
temperature $T_{coma} = 9$ keV. 
These solutions shown in Figure 8 
become essentially single phase 
($q \approx 0$) for $r \gta 5$ kpc. 
Beyond a few kpc the apparent temperature 
$T(R)$ exceeds the temperature at physical 
radius $r = R$, $T(r)$, because of line of sight projection.
The mean temperature in the two regions 
as in Equations (7) and (8) are:
$\langle T_1 \rangle = 1.0$ keV and
$\langle T_2 \rangle = 8.5$ keV.
$\langle T_2 \rangle$ is only 0.4 keV less than 
that observed by Vikhlinin et al., within the 
margin of observational error. 
The surface brightness $\Sigma_x$ (0.2 - 5) keV 
of this variable $q$ model shown 
in Figure 7 is also a better fit to the observations at all $R$.

If the dark and stellar mass of NGC 4874 are ignored as 
well as stellar mass loss and supernovae, we find that 
no galactic-scale 
cooling flow develops in our Coma model even after 
computing for  
$t_n - t_i = 13$ Gyrs when the central temperature 
has dropped by only $\sim 0.5$ keV below its initial
value.
Although there is no cluster-scale cooling flow in Coma, 
more relaxed clusters with $T_{vir} \sim 9$ 
have higher central densities and can develop cooling flows 
even without a central cD.

We have also performed NGC 4874/Coma calculations including thermal
conductivity as suggested by Vikhlinin et al. (2001).  
Conduction with
unmodified Spitzer and saturated coefficients can definitely be ruled
out since the gas temperature remains isothermal at $\sim 8$ keV
throughout NGC 4874 even for large $t_n - t_i$.  
Our results do not
support the suggestion of Vikhlinin et al. that 
the sharp drop in gas
temperature from 8 keV to 1 keV over a few kpc 
can be understood simply by setting the 
conductive heat flow (using a 
conductivity at the classical Spitzer level) 
at some radius equal to the interior radiative X-ray luminosity 
near NGC 4874.
Instead we find
that the coefficients for saturated and unsaturated conduction must be
reduced by more than 300 before $\langle T_1 \rangle$ drops below 1.5
keV.  
As the conductivity is reduced further, our solutions approach
the observed $\langle T_1 \rangle = 1$ keV 
just as the conductivity term in
the thermal energy equation becomes too small to influence the
temperature profile at all. 
This result for conduction in 
NGC 4874/Coma is essentially the same 
for either constant or variable $q$ flows. 
We conclude that thermal conduction in
NGC 4874/Coma is magnetically suppressed to levels of negligible
relevance.

In summary, {\it Chandra} and {\it XMM} 
temperature profiles near NGC 4874 can be fit reasonably well
with either constant or variable dropout $q$, 
although variable $q$ flows are in better agreement 
with {\it Chandra} observations.
As with M87/Virgo, the preference between these two 
models must be based on the evidence in 
high quality X-ray spectrum for or against multiphase flow.

\section{Variation of ${\rm {\dot M}}(r)$}

As we mentioned in the Introduction, 
the observed density and temperature profiles 
can be used to estimate $t_{cool}(r)$ and therefore 
the integrated mass flow ${\dot M}(r)$ at radius $r$ for 
steady state cooling flows. 
Typically, it is found that ${\dot M} \propto r$ 
for $r < r_b \sim 40 - 120$ kpc (e.g. Allen et al 2001).
The decomposition 
procedure often used by X-ray astronomers to find 
${\dot M}(r)$ from the observed $T(r)$ and $n(r)$ 
is based on dividing the observed cluster 
into an ensemble of concentric spherical shells 
and determining the bolometric X-ray emission 
from each shell, assuming steady state inflow with 
no stellar mass loss or heating 
(e.g. Arnaud 1988; Allen et al. 2001). 
The luminosity of the $j$th shell is
\begin{equation}
L_j = \Delta {\dot M}_j (H_j + \Delta \Phi_j)
+ \left[ \sum_{i=1}^{j-1} \Delta {\dot M}_i 
(H_j + \Delta \Phi_j) \right] 
\end{equation}
where $\Delta {\dot M}_j$ is the mass deposited in 
the $j$th shell, 
$H_j = 5 k T_j/ 2 \mu m_p$ is the enthalpy 
of the $j$th shell and 
$\Delta \Phi_j$ is the gravitational energy 
released in crossing the shell. 
The physical significance of each term is 
explained in Arnaud (1988).
It is interesting to compare the mass flow 
$${\dot M}_{12}(r_j) 
= \sum_{i = 1}^{j-1} \Delta {\dot M}_i$$
found from Equation (12) with 
${\dot M}(r)$ from our computed models.

The heavy solid lines in the upper two panels of 
Figure 9 show ${\dot M}(r) = \rho u 4 \pi r^2$ 
for our $q = 1$ and variable $q$ flows
in M87/Virgo at time $t_n = 13$ Gyrs.
Also shown are the rates 
${\dot M}_{do}(r) = \int_0^r 4 \pi r^2 
(q \rho / t_{cool}) dr$ that mass is 
dropping out of the flow (dotted lines); 
note that this corresponds to the integrated 
spectroscopic mass deposition rate. 
If the flow were perfectly steady state 
${\dot M}(r)$ and ${\dot M}_{do}(r)$ should be equal 
as they nearly are for the $q = 1$ solution.
For variable $q$ flow, however, $q \approx 0$ 
beyond $\sim 5 - 10$ kpc, so the integrated dropout mass  
levels off (dotted line in center panel).
For this flow  
${\dot M}(r) \approx {\dot M}_{do}(r)$ only 
holds for $r \lta 20$ kpc and beyond this radius 
they diverge in magnitude and sign 
because of small deviations from steady flow.
The approach to steady flow is accelerated 
in $q = 1$ flow (upper panel) 
because mass dropout tends to 
damp small transient velocities, i.e. dropout 
makes subsonic flow even more subsonic.

In Figure 9 we also illustrate
the mass deposition rate ${\dot M}_{12}(r)$ determined
by applying Equation (12) above 
to our computed flow at time $t_n$ (dashed lines).
In solving this equation, we used the same gravitational
potential used in our M87 flow calculations and
evaluated the enthalpy using the temperature $T_h(r)$
of the background flow at time $t_n$.
For $q = 1$ flow ${\dot M}_{12}(r)$ agrees quite 
well with both ${\dot M}(r)$ (solid line) 
and ${\dot M}_{do}(r)$ (dotted line) 
in spite of the somewhat transient nature of the flow
and the presence of stellar mass loss in 
$r \lta 10$ kpc.
However, Equation (12) fails to capture
either the mass flow or
mass deposition profiles in the variable $q$ flow 
(Fig. 9 center panel).
Even when Equation (12) is inaccurate, we seem to find 
the classic result, ${\dot M}_{12} \propto r$.
In the variable $q$ solution beyond about 10 kpc,  
small undamped radial motions remain at time $t_n$ 
which have almost no influence on $n(r)$ or $T(r)$, 
but which radically alter ${\dot M}(r)$. 
We conclude that Equation (12) is not a reliable 
means of estimating ${\dot M}(r)$ in clusters 
that have small deviations from steady flow. 
Indeed, this may apply to all clusters observed.

For comparison we also plot (lower panel of
Figure 9) ${\dot M}(r)$ and ${\dot M}_{do}(r)$
based on a variable $q$ calculation
in which the interstellar flow in
M87 is computed as an isolated galaxy
(as in Brighenti \& Mathews 1999), then inserted into
Virgo at time $t_i = 6$ Gyrs, i.e. the ``alternative''
evolutionary model previously discussed.
This computational procedure introduces fewer 
velocity transients. 
The resulting model generates $n(r)$ and $T(r)$
profiles at time $t_n$ that are virtually identical to 
those plotted in Figure 2, but the corresponding 
${\dot M}(r)$, shown as a solid line in  
the lower panel of Figure 9, is radically 
different from ${\dot M}(r)$ in the center panel.
${\dot M}(r)$ in the alternative solution 
(solid line) lies closer to 
${\dot M}_{12}(r)$ (dashed line) but both deviate from 
the true mass dropout profile ${\dot M}_{do}(r)$ 
(dotted line).
Asymptotically for large $t_n - t_i$, 
we expect both ${\dot M}(r)$ 
and ${\dot M}_{12}(r)$ to approach ${\dot M}_{do}(r)$, 
but this may take longer than a Hubble time.
The approximate agreement between ${\dot M}(r)$ and 
${\dot M}_{12}(r)$ estimated from Equation (12) 
in $r \lta 50$ may be fortuitous. 
Equation (12) is nonetheless incapable of 
accurately detecting 
the presence of radially dependent $q(r)$ in our models. 

Finally, in Figure 10 we show ${\dot M}$, 
${\dot M}_{do}$ and ${\dot M}_{12}$ for the two models 
with thermal conductivity suppression factors $f$ 
chosen to best fit the 
$n(r)$ and $T(r)$ data in Figure 5.
Both of these are alternative models in which the flow 
in M87 is computed before inserting into Virgo.
When $q = 1$ the conductivity has little effect on 
the various mass flow rates.
But flows with $q = {\rm exp}(-0.05/n)$ show that 
${\dot M}_{do}$ can be reduced by about 3 from the 
corresponding calculation without thermal conduction
(lower panel in Figure 9). 
As before, Equation (12) is inaccurate when applied 
to solutions with variable $q$ and continues to indicate 
${\dot M}_{12} \propto r$, predicting mass deposition 
rates that are much too large.

\section{Conclusions}

Recent {\it Chandra} and {\it XMM} observations of M87 in the 
Virgo cluster and NGC 4874 in Coma show that the gas temperature 
decreases dramatically from the cluster virial temperature 
(3 or 9 keV respectively) at $\sim 50$ kpc 
down to the galactic virial temperature 
($\sim 1$ keV) on scales of a few kiloparsecs.
We have performed a series of gas dynamical calculations to 
determine the physical nature of this sharp thermal gradient. 

In addition, recent observations have presented somewhat 
contradictory information about the important question of 
single phase or multiphase cooling flows.
In multiphase flows some of the hot gas cools 
throughout a large volume to avoid accumulation of large masses 
of cooled gas right at the center of the flow. 
Multiphase flow can be identified from spectral observations 
that show a large range of temperatures at a particular 
projected radius in the flow, larger than that 
expected from radial temperature variations projected 
along the line of sight. 
As a result of local cooling, gas mass ``drops out'' from 
multiphase flows and the gas 
density in the flow decreases as 
$(\partial \rho / \partial t)_{do} = - q \rho / t_{cool}$ 
where $t_{cool}$ is the local radiative
cooling time and $0 \le q(r) \lta 1$ is a mass dropout 
function that is chosen for a best fit to the observations.

Our gas dynamical studies of the interaction of cluster 
and interstellar gas in cluster-centered cD galaxies are necessarily 
approximate because of unknown stochastic merging events that 
may have influenced the cD environment in the past. 
Nevertheless, it is possible to estimate the flow of cluster 
gas into central cD galaxies provided the present day 
properties of the cluster gas have remained approximately constant 
over times for the flows to become quasi-steady on galactic scales.
In most of the gas (cooling) 
flows calculated here we abruptly place the cD galaxy into 
the cluster at some time $t_i$ after which there is 
a transient flow that relaxes to the desired solution 
after several Gyrs. 
We have also performed ``alternative'' evolutionary  
calculations in which the central cD 
galaxy evolves by itself from early times, similar to our 
galaxy group model for 
NGC 4472 (Brighenti \& Mathews 1999), 
and is placed in the cluster environment at time $t_i$. 
The results of these two types of flow 
calculations are very similar 
except the total mass of cooled gas is
somewhat larger in the alternative flows.

As a result of our calculations for M87/Virgo and NGC4874/Coma,
we reach the following conclusions:

\vskip.2cm
\noindent
(1) The radial variation of gas temperature and density on 
scales of the cD galaxy can be understood from the usual 
cooling flow assumptions (e.g. stellar mass loss, spatially 
distributed mass dropout, etc.) without requiring additional heating 
sources such as thermal conduction.

\vskip.2cm
\noindent
(2) If we perform our gas dynamical calculations 
for the Virgo cluster, assuming isothermal 
gas at $T \sim 3$ keV with the currently observed 
large scale density distribution, 
we find that a fully mature, large scale 
cooling flow does not develop 
without the presence of the central cD galaxy.
After many Gyrs of calculation, a cD-less single phase flow
in the Virgo potential exhibits line of sight
averaged gas temperature profiles $T(R)$ with
a broad central minimum, but no gas cools.
However, a fully complete cosmological calculation of 
the Virgo cluster is expected to produce a cooling flow 
with or without a central galaxy 
(e.g. Suginohara \& Ostriker 1998).

\vskip.2cm
\noindent
(3) In our model for 
the richer Coma cluster with virial gas temperature 
$\sim 9$ keV, we find that the central density is sufficiently 
low that no cluster scale 
cooling flow develops in a Hubble time  
without the central cD. 
In the presence of the cD galaxy, a cooling flow forms on 
a galactic scale, just a few kpc in size. 
This conclusion is probably not general. 
Since Coma is dynamically disturbed,
its current central cluster gas density is atypically low. 
Relaxed clusters of similar richness have higher 
central gas densities and may form global 
cooling flows even without a central cD galaxy.

\vskip.2cm
\noindent
(4) The radial gas density and temperature profiles in both 
M87/Virgo or NGC4874/Coma can be understood either in terms 
of models with large constant mass dropout at every radius 
($q = 1$; multiphase flow), 
or with a variable dropout coefficient $q(r)$ that 
becomes very small (single phase flow) beyond $\sim 10$ kpc. 
High quality X-ray spectra can detect the full range of 
local temperatures in the flow and thus determine the 
single or multiphase character of the flow. 
An observation of the oxygen ratio 
OVII(0.574 keV)/OVIII(0.652 keV) would be useful 
to detect gas cooling below $10^6$ K. 
However, currently available 
{\it XMM} observations of M87 have been unable to 
detect unambiguous evidence for multiphase cooling, 
even in the central regions of M87. 

\vskip.2cm
\noindent
(5) If we insist that the gas must cool somewhere in the 
flow, we suggest that this occurs within the optical 
extent of the central cD galaxy where the gas temperature 
is close to the galactic virial temperature $\sim 1$ keV.
At such low temperatures it is observationally more 
difficult to detect intrinsic multiphase flow than  
in hotter gas more distant from the cD. 
These remarks are specific to NGC4874/Coma and M87/Virgo; 
in more massive cooling flow clusters, 
the cooled dropout mass may 
be distributed over a larger spatial region.

\vskip.2cm
\noindent
(6) Unfortunately, when most of the radiative 
cooling is restricted to the central regions of the cD galaxy
($r \lta 10$ kpc), the mass of cooled gas there  
may be inconsistent with 
mass to light ratios determined from stellar kinematics. 
At early times when most of the cooling mass dropout occurred,
the cooled gas probably formed into luminous stars. 
Stellar H$\beta$ observations suggest 
that luminous stars may continue to form from cooling dropout.
If the dropout mass is nonluminous, as many authors assume, 
then it is easy to 
violate observations of the stellar mass to light ratio 
unless the nonluminous mass is 
widely dispersed over the inner $\sim 7$ kpc in M87.

\vskip.2cm
\noindent
(7) In this paper we have sought ways to hide the
multiphase nature of gas cooling within 10 kpc of the
center of M87 and to hide the presence of large masses
of optically dark cooled gas there.
In spite or our limited success,
it is difficult to fully embrace this approach.
If M87 were in a much richer cooling flow cluster,
it may be impossible to assert that all of the
cooled gas accumulates within the central 10 kpc.

\vskip.2cm
\noindent
(8) We confirm early studies that the coefficient for 
thermal conductivity must be unrealistically 
fine-tuned or entirely negligible to lower 
the mass deposition rate without disagreeing with the observed 
temperature profiles.

\vskip.2cm
\noindent
(9) For the cooling flow models considered here we find that 
the mass flow rate ${\dot M}(r) = \rho u 4 \pi r^2$ can be 
very sensitive to
small transient velocities in the flow that
have no consequence for $n(r)$ and $T(r)$.
Furthermore, the often used Equation (12) is incapable of 
accurately detecting the presence of radially dependent 
mass dropout $q(r)$ in flows that are not completely 
relaxed, which may apply to the majority of cooling flows. 
In particular, when flows become nearly single 
phase at large radius, as indicated by recent {\it XMM} data, 
our models suggest that ${\dot M}_{12}(r)$ found from 
Equation (12) predicts a larger mass dropout at the 
current time than really exists in the flow.

%\end{document}

\vskip 1in

\vskip.4in
Studies of the evolution of hot gas in elliptical galaxies
at UC Santa Cruz are supported by
NASA grant NAG 5-3060 and NSF grant
AST-9802994 for which we are very grateful.
FB is supported in part by grants MURST-Cofin 00
and ASI-ARS99-74.

%\end{document}

\clearpage

\vskip.1in
\figcaption[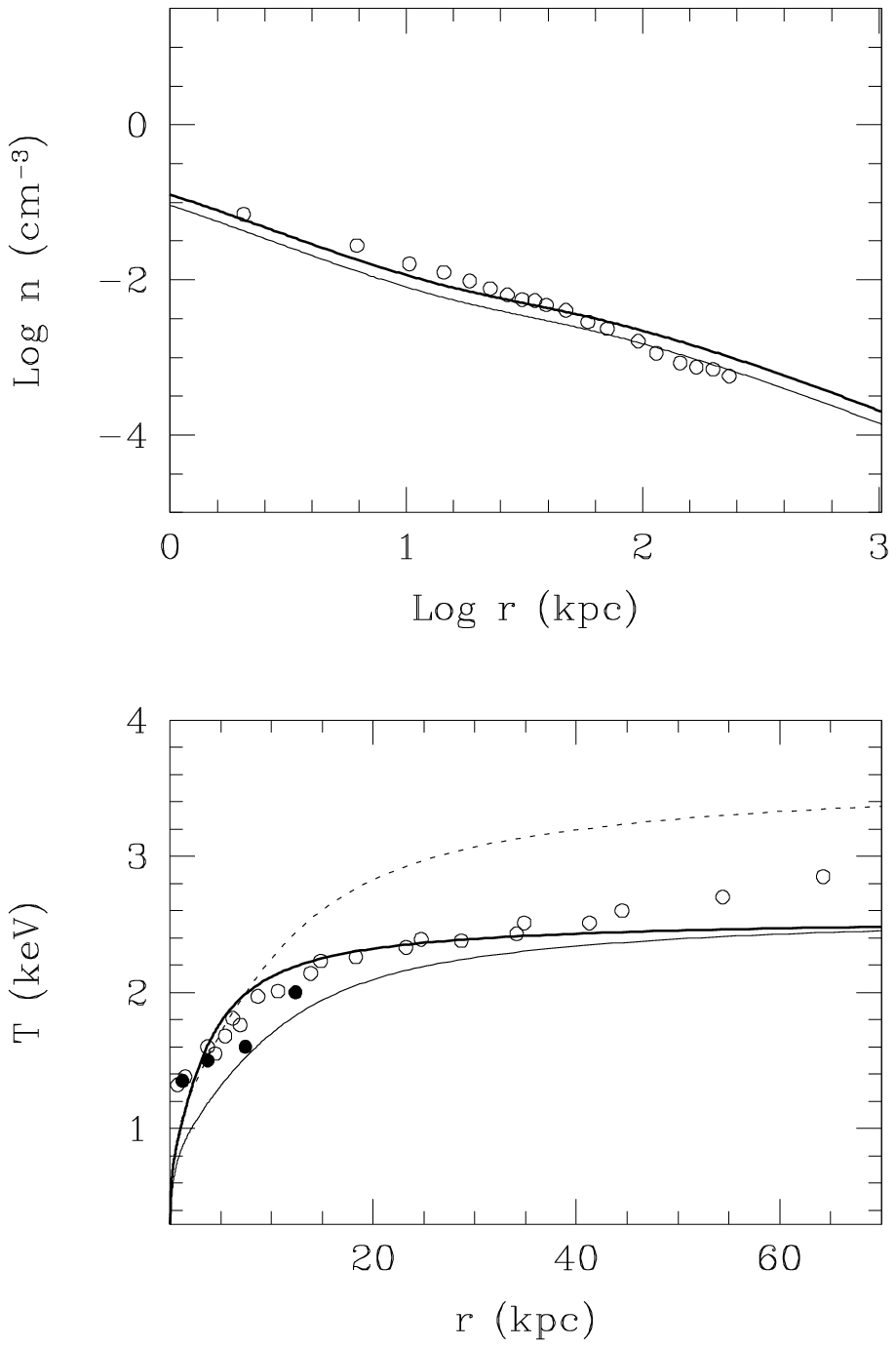]{
Gas density and temperature for  
the $q = 1$ model of M87/Virgo at time $t_n$.
{\it Upper panel}: Density profiles $n(r)$.  
{\it Light solid line}: density of the background hot gas,
{\it heavy solid line}: mean local density including emission 
from cooling regions.
The data are taken from Nulsen \& B\"ohringer (1995).
{\it Lower panel}: Temperature profiles for the $q = 1$ model.
{\it dotted line}: $T_h(r)$, the background, uncooled gas;
{\it light solid line}: $T(r)$, the mean emission-weighted 
local temperature at physical radius $r$ including cooling regions; 
{\it heavy solid line}: $T(R)$ emission-weighted mean of 
$T(r)$ along the line of sight at projected radius $R = r$.
Data are from Molendi \& Gastaldello (2001) ({\it filled circles})
and B\"ohringer et al. (2001) ({\it open circles}).
\label{fig1}}

\vskip.1in
\figcaption[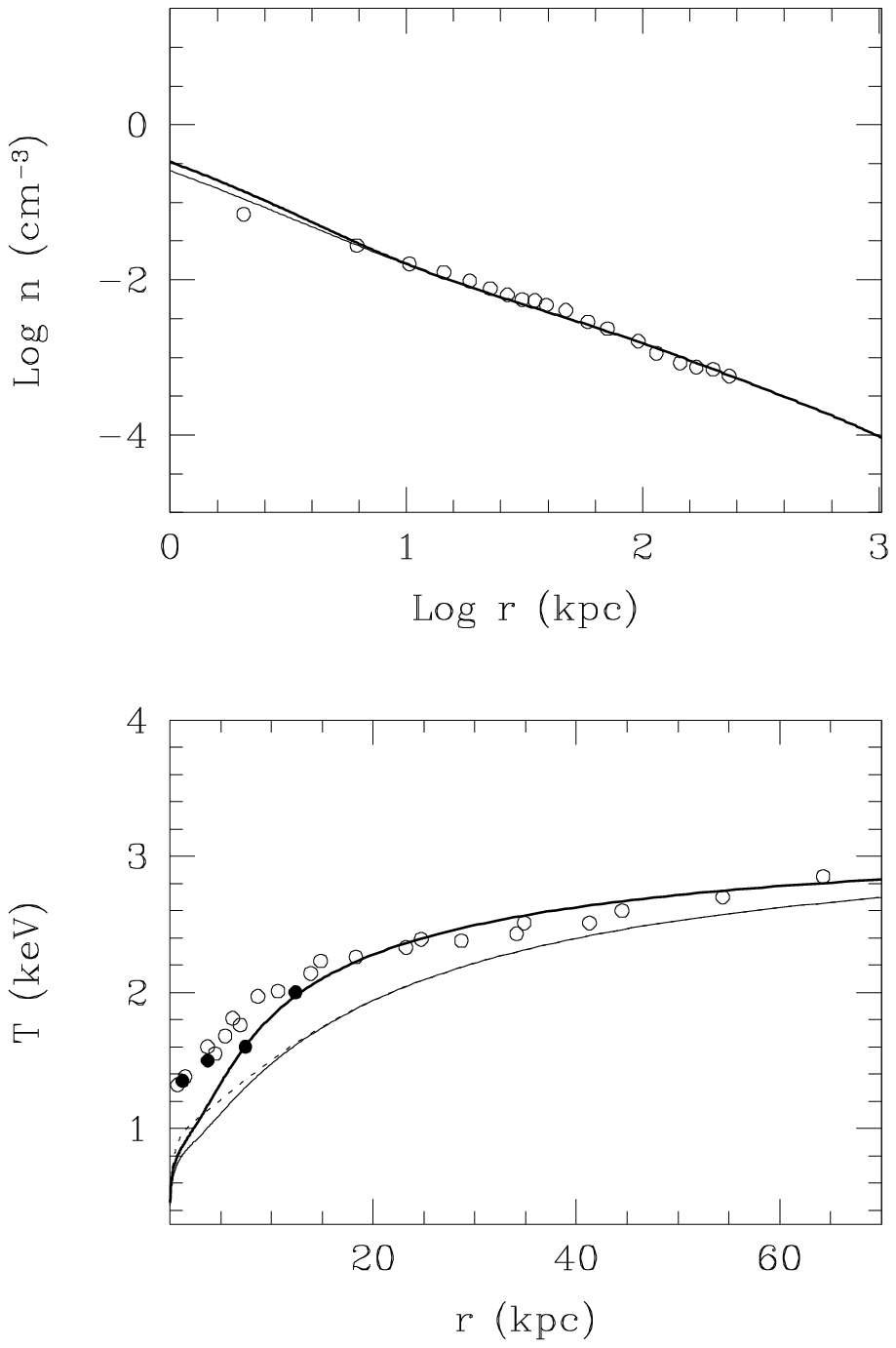]{
Gas density and temperature for 
the variable $q$ model of M87/Virgo at time $t_n$.
Line types and data are identical to those in Figure 1.
\label{fig2}}

\vskip.1in
\figcaption[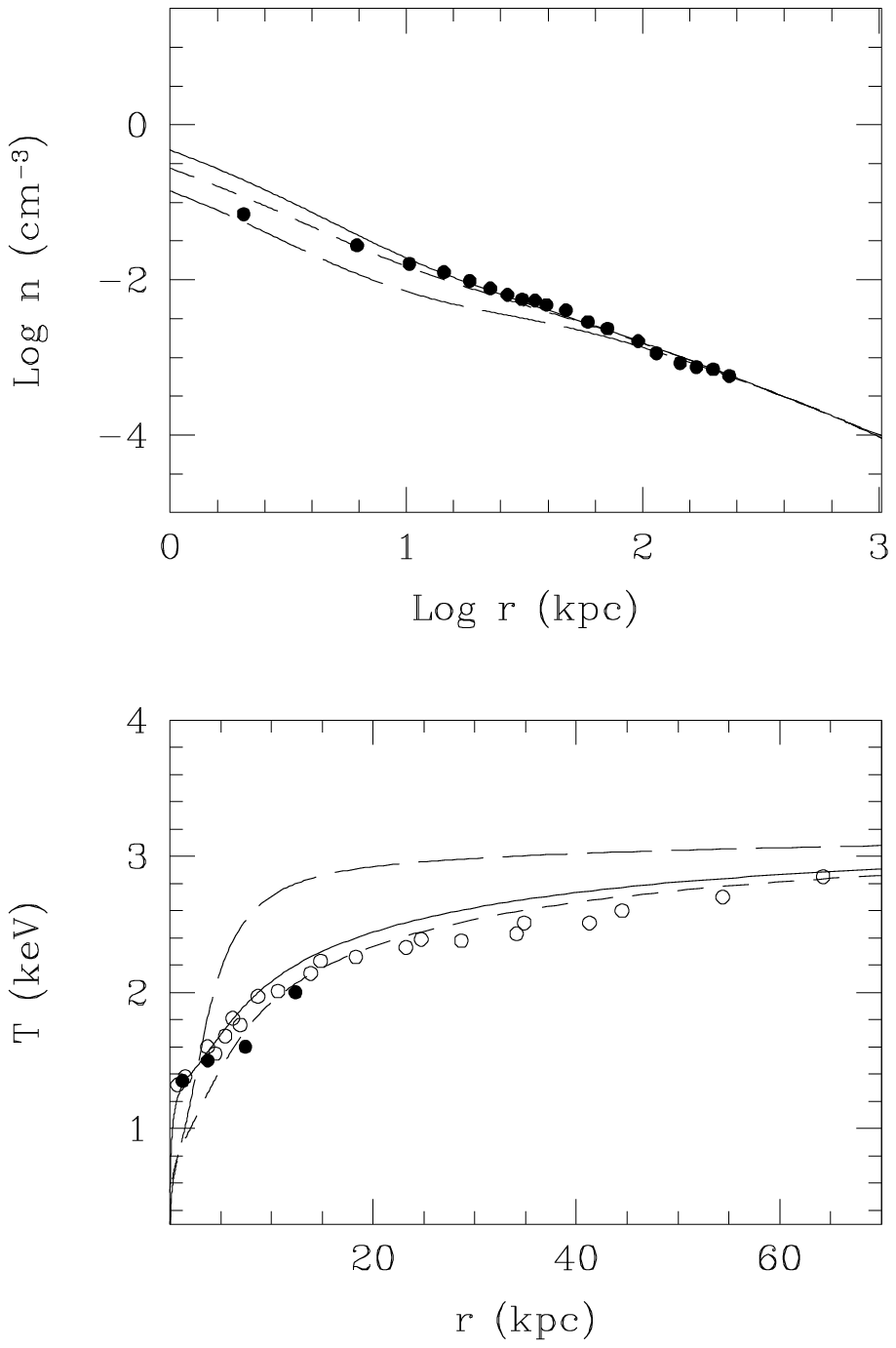]{
Gas density and temperature in M87/Virgo at time $t_n$ 
for three variants of the variable $q$ model.
In the lower panel only the projected mean 
temperature $T(R)$ is plotted. 
Data are identical to that in Fig. 1 .
{\it Long dashed lines}: Flow with a more recent initial 
time $t_i = 10$ Gyrs;
{\it Solid lines}: Flow with stellar $M/L_B$ increased 
to 16;
{\it Short dashed lines}: Flow with $\alpha_*$ reduced 
by 2 with $M/L_B = 8$.
\label{fig3}}

\vskip.1in
\figcaption[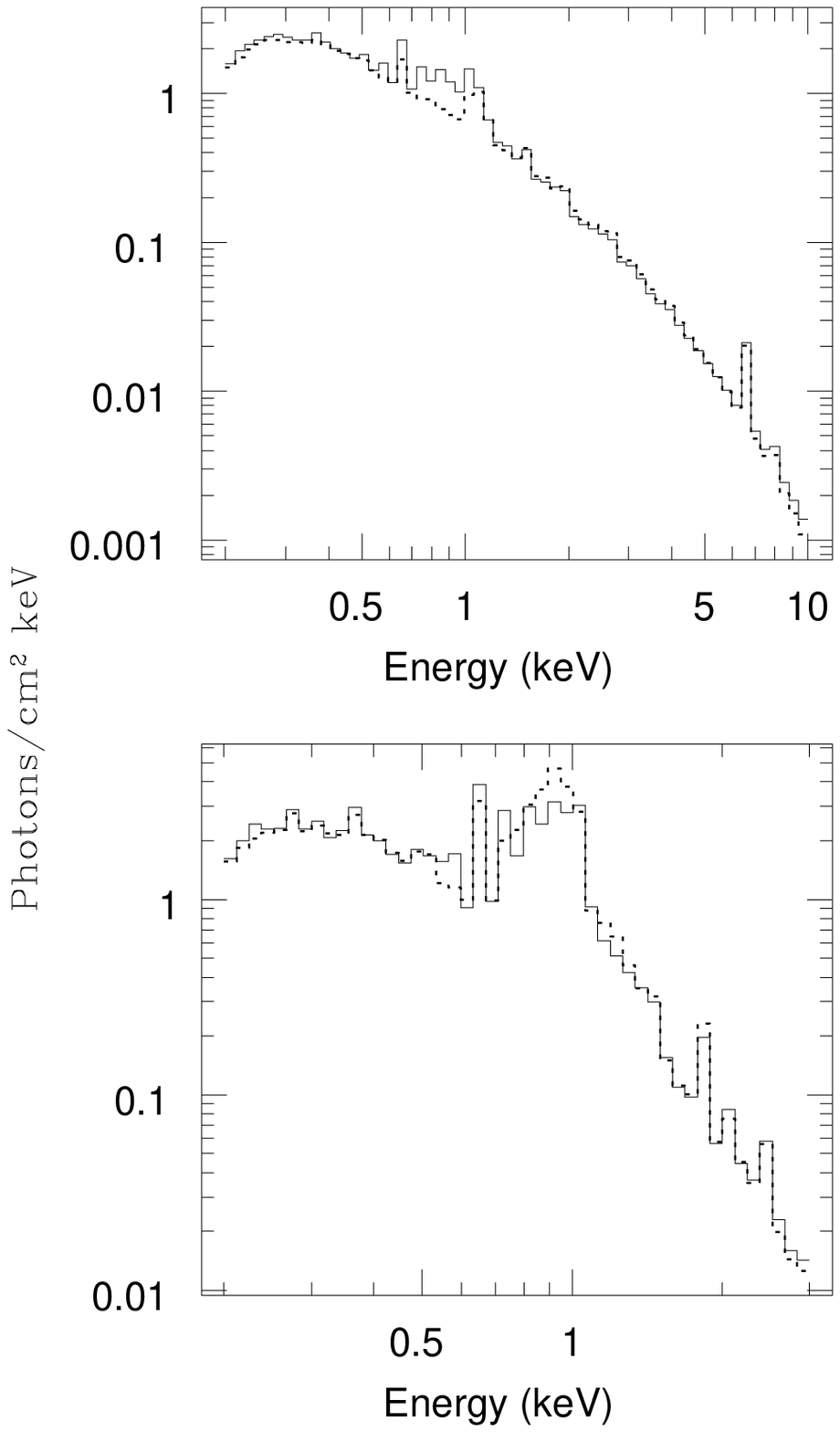]{
A comparison of single temperature (1T) spectra
(for variable $q$ M87 model)
with a composite spectra of a  
single temperature plus cooling flow (1T + CF) 
with equal normalizations 
(for $q = 1$ M87 model). The CF spectrum extends 
from $T_h$ down to 0.1 keV.
{\it Upper panel}: Spectra at 
$R \approx 30$ kpc: (1T + CF) composite spectrum for gas at 
$T_h = 3$ keV (solid line); 1T spectrum for gas at 
average temperature $T = 2.5$ keV (dotted line).
{\it Lower panel}:
$R \approx 1$ kpc: (1T + CF) composite spectrum for gas at
$T_h = 0.95$ keV (solid line); 1T spectrum for gas at
$T = 0.8$ keV (dotted line).
\label{fig4}}

\vskip.1in
\figcaption[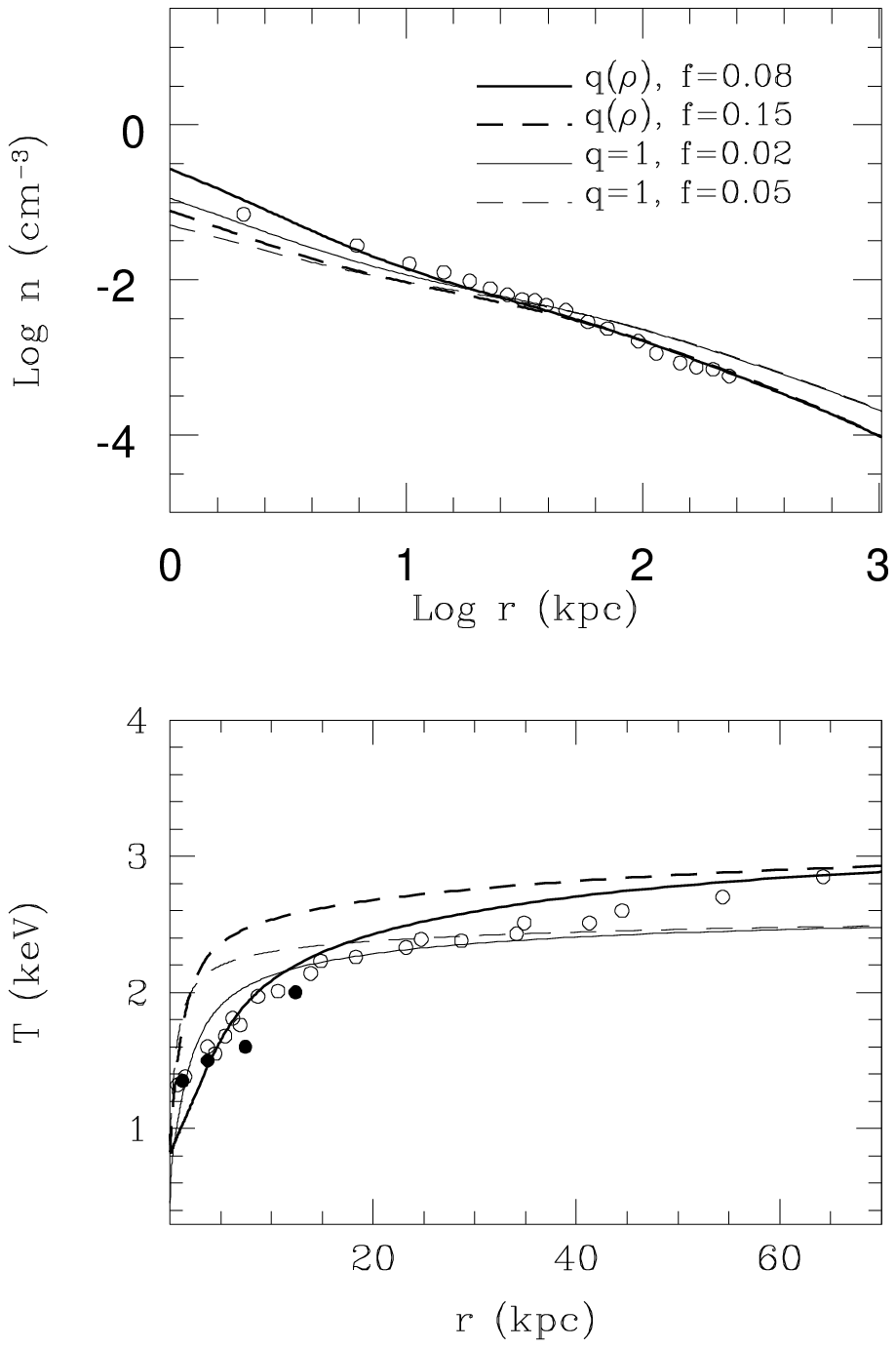]{
Gas density and temperature profiles for M87/Virgo 
at time $t_n$ including thermal conductivity reduced
at all radii  
from the Spitzer or saturated value by a factor $f$.
{\it Upper panel}: The effective density profile 
$n(r)$ corresponding to emission from background and 
cooling gas. 
Variable $q$ flow for $f = 0.02$: 
({\it heavy solid line}) and 
$f = 0.05$ ({\it heavy dashed line});
Flow with $q = 1$ for $f = 0.08$: 
({\it light solid line}) and 
$f = 0.15$ ({\it light dashed line});
{\it Lower panel}: Projected temperature $T(R)$ 
profiles; line types are identical to those in the 
upper panel.
\label{fig5}}

\vskip.1in
\figcaption[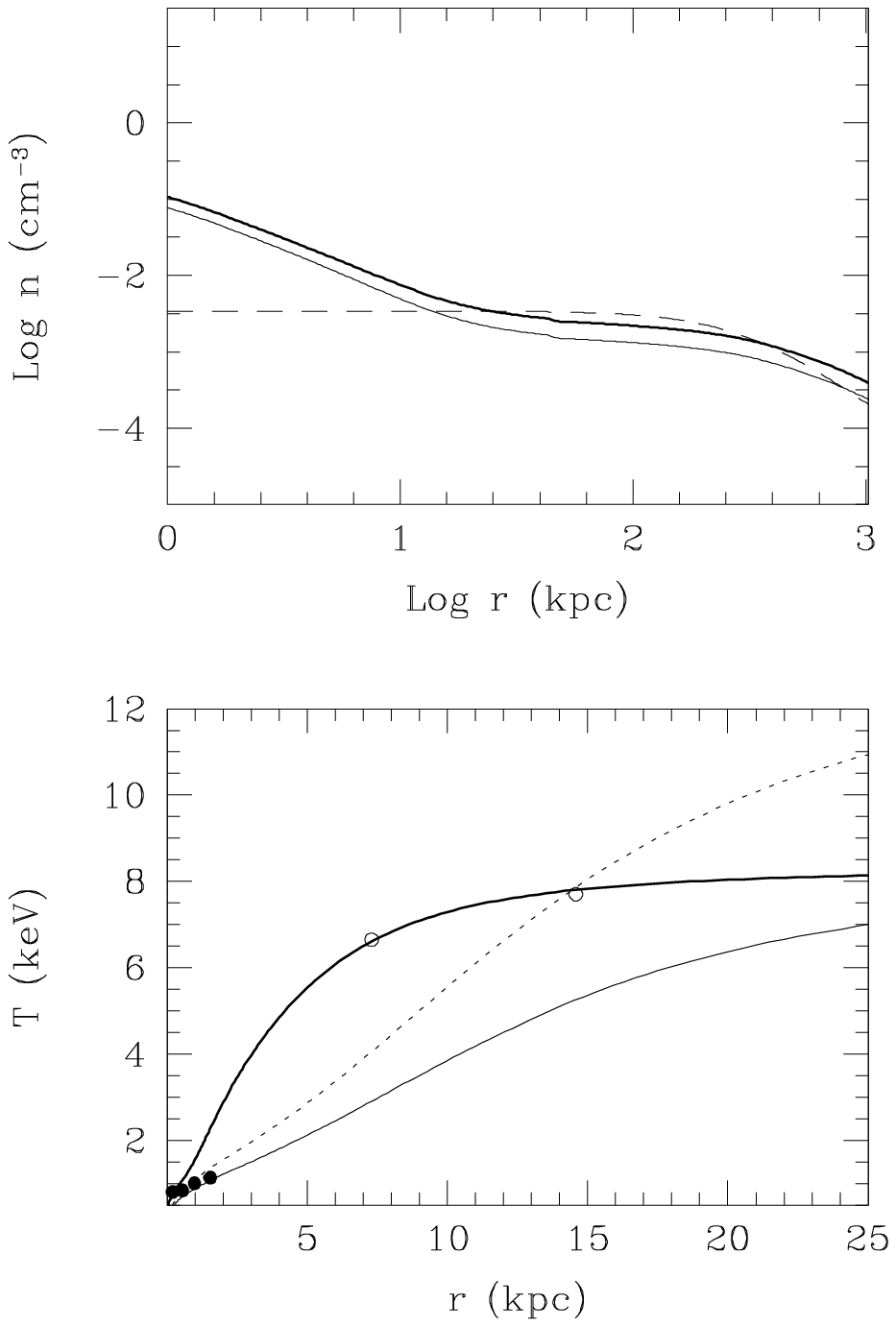]{
Gas density and temperature in 
the $q = 1$ model for NGC4874/Coma at time $t_n$.
{\it Upper panel}: Density profiles $n(r)$ with 
solid curves labeled as in Fig. 1.
{\it Long dashed line}: $\beta$-model gas profile from 
Briel et al. (1992) appropriate for large $r$.
{\it Lower panel}: Temperature profiles for $T_h(r)$, 
$T(r)$ and $T(R)$ labeled as in Fig. 1.
Data are from Arnaud et al. (2001) ({\it open circles})
and Vikhlinin et al. (2001) ({\it filled circles}).
\label{fig6}}

\vskip.1in
\figcaption[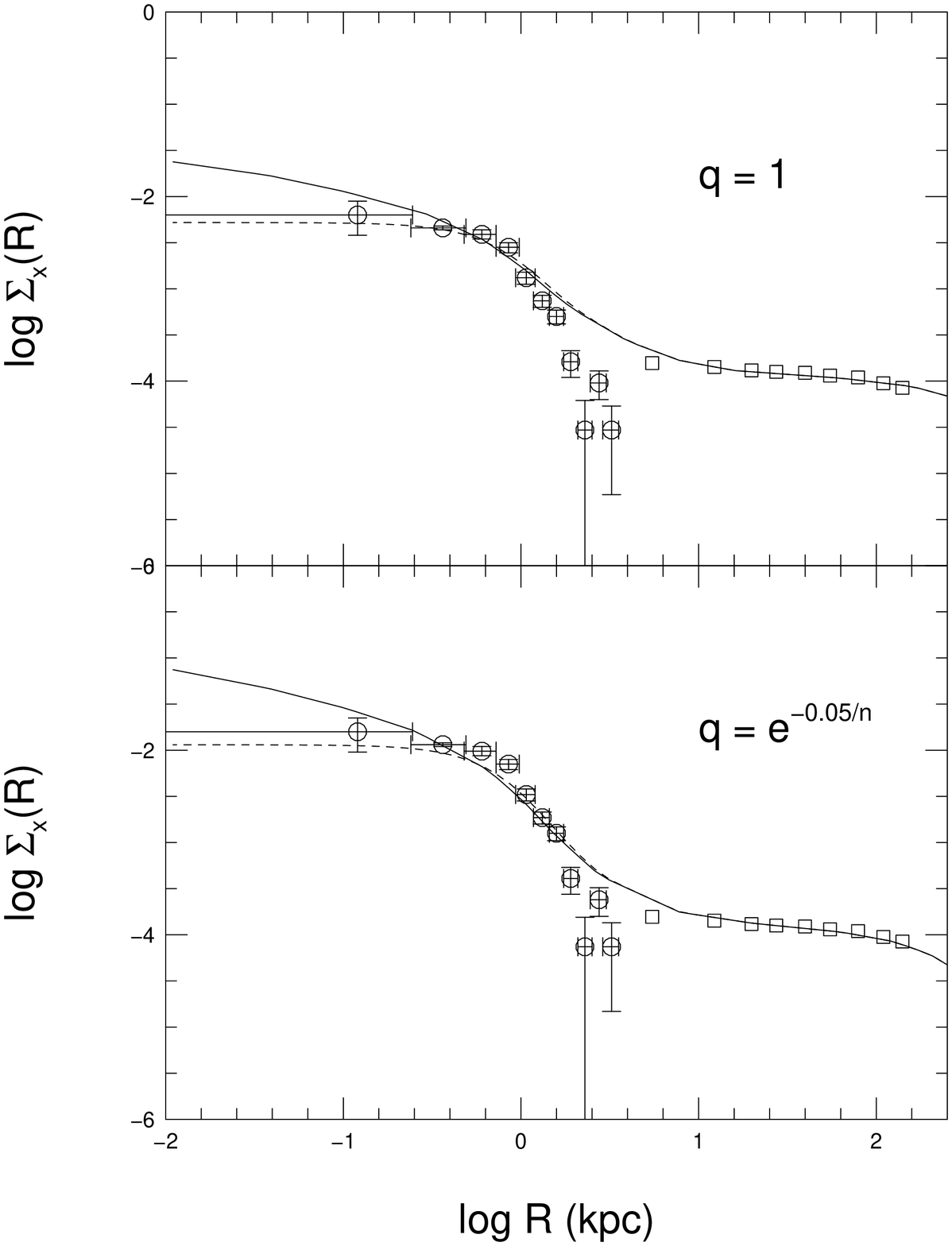]{
X-ray surface brightness distribution $\Sigma_x(R)$ 
(0.5 - 2 keV) at $t_n$ in NGC4784/Coma before 
({\it solid lines}) and 
after convolution with {\it Chandra} PSF ({\it dashed lines}).
{\it Upper panel}: For $q = 1$ flow;
{\it Lower panel}: For variable $q$ flow.
Circles are {\it Chandra} data from Vikhlinin et al. (2001) and 
squares are {\it ROSAT} data from Dow \& White (1995).
\label{fig7}}

\vskip.1in
\figcaption[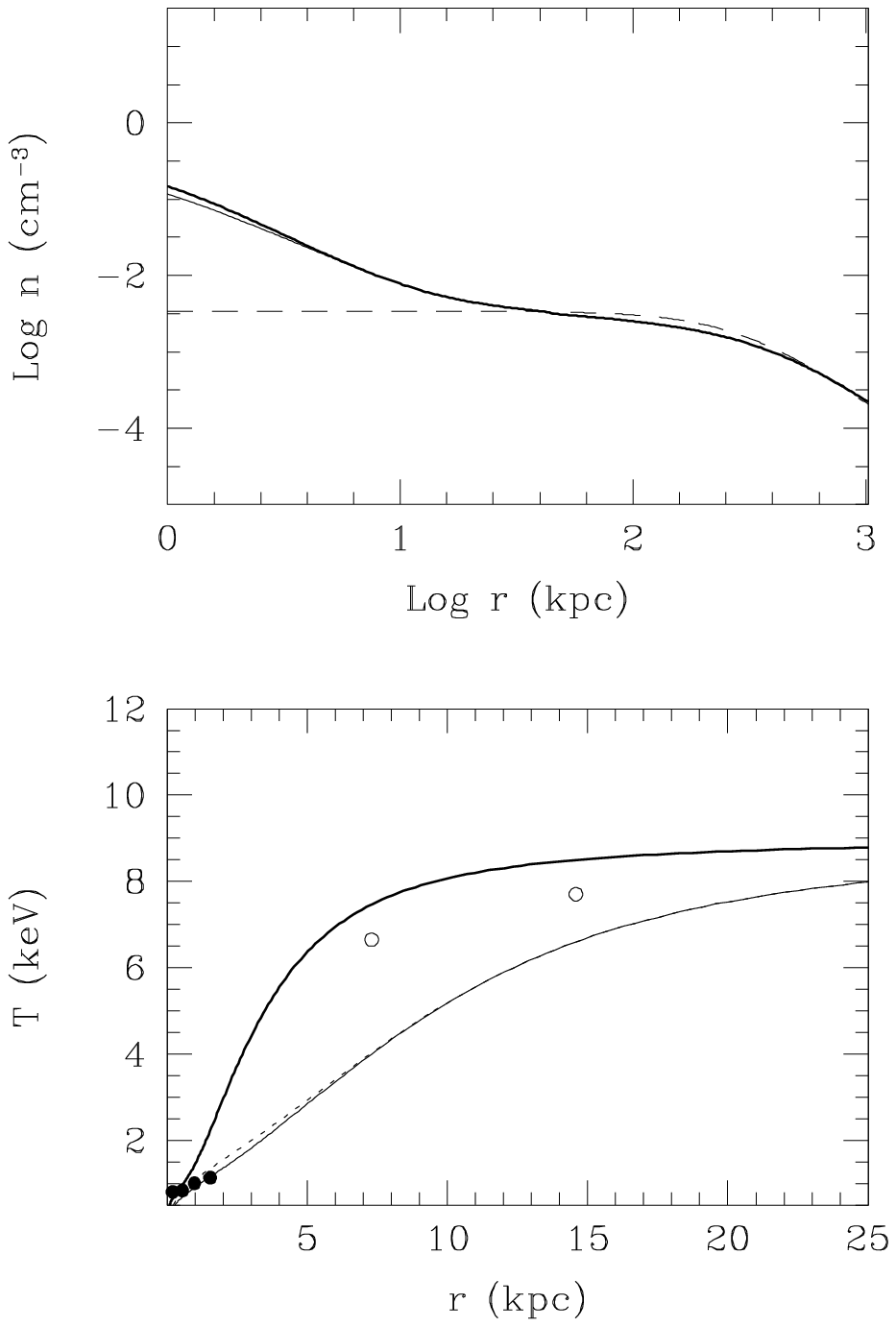]{
Gas density and temperature in 
the variable $q$ model for NGC4874/Coma at time $t_n$.
Line types and data are identical to those in Figure 6.
\label{fig8}}

\vskip.1in
\figcaption[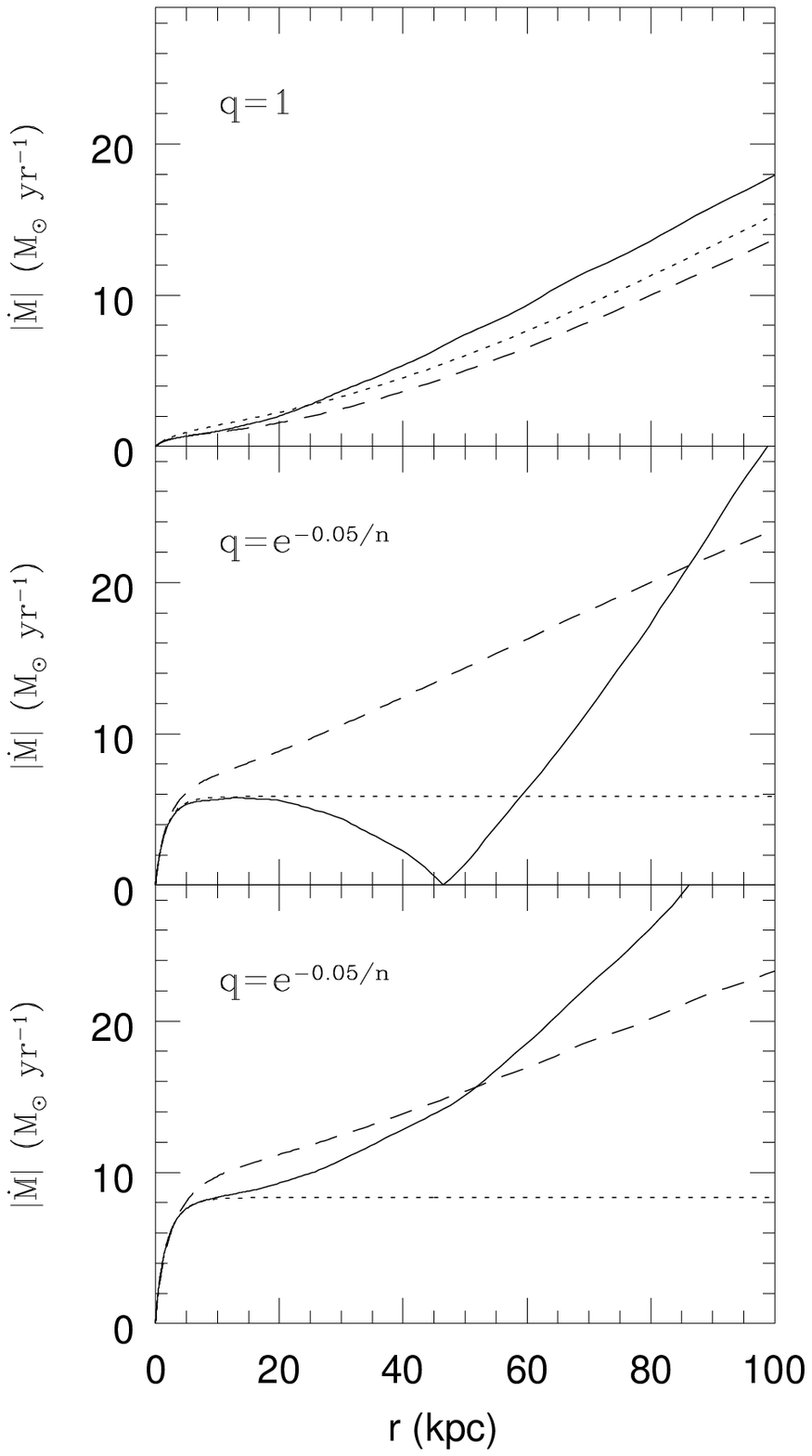]{
Mass flows in M87. 
Variation of ${\dot M}(r) = \rho u 4 \pi r^2$ 
in computed models ({\it solid lines}),  
${\dot M}_{12}(r)$ from Equation (12) ({\it dashed lines}) 
and integrated rate of cooling dropout ${\dot M}_{do}(r)$
({\it dotted lines}), all evaluated at time $t_n = 13$ Gyrs.
{\it Upper panel}: For the $q = 1$ solution.
{\it Center panel}: For the variable $q$ solution. 
${\dot M}(r)$ computed from the model (solid line) 
changes sign and becomes positive (outflow) 
for $r > 53$ kpc. 
{\it Lower panel}: An alternative variable $q$ solution 
at $t_n$ in which M87 is evolved separately, then 
inserted into Virgo at $t_i = 6$ Gyrs. 
\label{fig9}}

\vskip.1in
\figcaption[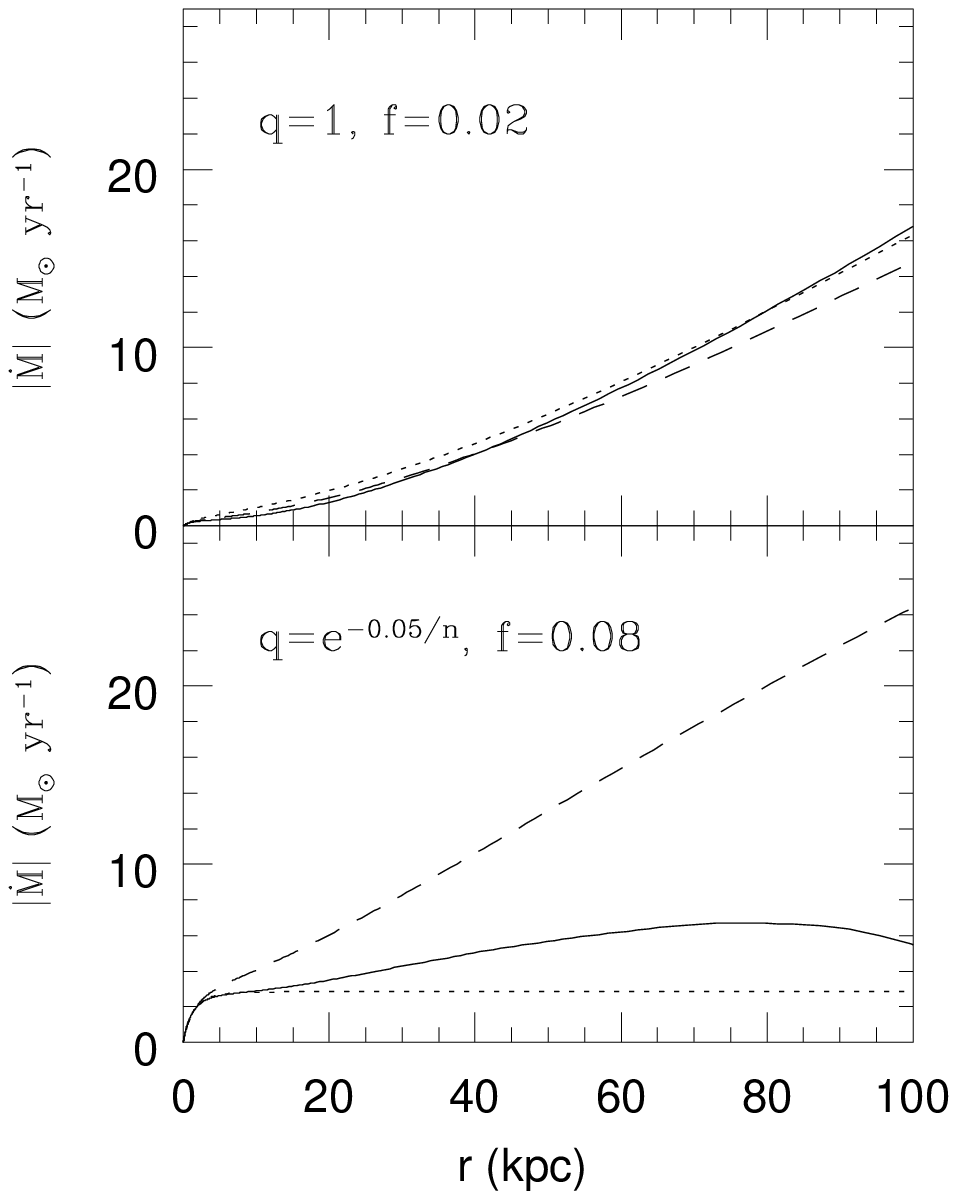]{
Radial variation of ${\dot M}(r)$ ({\it solid lines}), 
${\dot M}_{do}(r)$ ({\it dotted lines}) and 
${\dot M}_{12}(r)$ ({\it dashed lines}) for 
flows with thermal conductivity reduced by a factor $f$.
{\it Upper panel}: $q =1$ flow with $f = 0.02$;
{\it Lower panel}: variable $q$ flow with $f = 0.08$.
\label{fig10}}

\end{document}